\documentclass[fleqn,usenatbib]{mnras}

\makeatletter
\newcommand{\ifmnras}[2]{\@ifclassloaded{mnras}{#1}{#2}}
\makeatother
\usepackage{microtype}
\usepackage[T1]{fontenc}
\usepackage[utf8]{inputenc}
\usepackage{multicol}
\usepackage{newtxtext,newtxmath}
\usepackage{ae,aecompl}
\usepackage{fix-cm}

\usepackage[english]{babel}

\usepackage[svgnames]{xcolor}
\definecolor{equationCol}{HTML}{D95F02}
\definecolor{sectionCol}{HTML}{7570B3}
\definecolor{citeCol}{HTML}{1B9E77}

\ifmnras{\usepackage{hyperref}}{\usepackage[colorlinks=true, linkcolor=sectionCol, citecolor=citeCol]{hyperref}}

\ifmnras{}{\usepackage[a4paper, width=200mm, top=25mm, bottom=25mm]{geometry}}

\usepackage{fancyhdr}
\usepackage{titlesec}
\usepackage{cleveref}

\crefformat{footnote}{#2\footnotemark[#1]#3}

\usepackage{mathtools}
\usepackage{amsmath}
\usepackage{nicefrac}
\usepackage{breqn}
\usepackage{gensymb}

\usepackage{array}
\usepackage{booktabs}
\usepackage{arydshln}

\usepackage{subcaption}
\usepackage{graphicx}
\usepackage{float}

\usepackage{siunitx}

\ifmnras{}{\usepackage[natbibapa, nodoi]{apacite}}

\ifmnras{}{\usepackage{natbib}}

\usepackage{booktabs}

\ifmnras{}{\newcommand{\apj}{ApJ}}

\ifmnras{}{\newcommand{\aj}{AJ}}

\ifmnras{}{\newcommand{\apjl}{ApJL}}

\ifmnras{}{\newcommand{\mnras}{MNRAS}}

\ifmnras{}{\newcommand{\pasp}{PASP}}

\ifmnras{}{\newcommand{\nat}{Nature}}

\ifmnras{}{\newcommand{\aap}{AAP}}

\ifmnras{}{}

\newcommand*{\SavedEqref}{}
\let\SavedEqref\eqref
\renewcommand*{\eqref}[1]{%
    \begingroup
        \hyperseup{
            linkcolor=equationCol,
            linkbordercolor=equationCol,
        }%
        \SavedEqref{#1}%
    \endgroup
}

\newcommand*\lbreak{\\[\baselineskip]}

\newcommand{\scc}{SCL}

\newcommand{\kep}{\textit{Kepler}/\textit{K2}}
\graphicspath{ {images/} }

\title[SN2017jgh Shock Cooling Lightcurve]{SN2017jgh - A high-cadence complete shock cooling lightcurve of a SN~IIb with the {\it Kepler} telescope}
\author[P. Armstrong et al.]{P. Armstrong,$^{1}$\thanks{Email: \href{mailto:patrick.armstrong@anu.edu.au}{patrick.armstrong@anu.edu.au}}
B. E. Tucker,$^{1,2,3}$
A. Rest,$^{4,5}$
R. Ridden-Harper,$^{4,5}$
Y. Zenati,$^{5}$
A. L. Piro,$^{6}$\and
S. Hinton,$^{7}$
C. Lidman,$^{1,56}$
S. Margheim,$^{9}$
G. Narayan,$^{10,11}$
E. Shaya,$^{12}$
P. Garnavich,$^{24}$\and
D. Kasen,$^{31,32,33}$
V. Villar,$^{47}$
A. Zenteno,$^{50}$
I. Arcavi,$^{13,14}$
M. Drout,$^{6,15}$
R.~J.~Foley,$^{16}$\and
J. Wheeler,$^{17}$
J. Anais,$^{18}$
A. Campillay,$^{19}$
D. Coulter,$^{16}$
G. Dimitriadis,$^{16}$
D. Jones,$^{16}$\and
C. D. Kilpatrick,$^{21,22}$
N. Mu\~{n}oz-Elgueta,$^{20}$
C. Rojas-Bravo,$^{16}$
J. Vargas-Gonz\'{a}lez,$^{23}$\and
J. Bulger,$^{25}$
K. Chambers,$^{25}$
M. Huber,$^{25}$
T. Lowe,$^{25}$
E. Magnier,$^{25}$
B. J. Shappee,$^{25}$\and
S. Smartt,$^{30}$
K. W. Smith,$^{30}$
T. Barclay,$^{26,27}$
G. Barentsen,$^{28}$
J. Dotson,$^{29}$
M. Gully-Santiago,$^{17}$\and
C. Hedges,$^{28,29}$
S. Howell,$^{29}$
A. Cody,$^{29}$
K. Auchettl,$^{3,16,57}$
A. B\'{o}di,$^{34,35}$
Zs. Bogn\'{a}r,$^{34,35}$\and
J. Brimacombe,$^{36}$
P. Brown,$^{51}$
B. Cseh,$^{34}$
L. Galbany,$^{37}$
D. Hiramatsu,$^{38,39}$
T. W.-S. Holoien,$^{6}$\and
D. A. Howell,$^{38,39}$
S. W. Jha,$^{55}$
R. K\"{o}nyves-T\'{o}th,$^{34,40}$
L. Kriskovics,$^{34}$
C. McCully,$^{38,39}$
P. Milne,$^{45}$\and
J. Mu\~{n}oz,$^{41,42}$
Y. Pan,$^{43}$
A. P\'{a}l,$^{34}$
H. Sai,$^{44}$
K. S\'{a}rneczky,$^{34}$
N. Smith,$^{45}$
\'{A}. S\'{o}dor,$^{34,35}$\and
R. Szab\'{o},$^{34,35,46}$
R. Szak\'{a}ts,$^{34}$
S. Valenti,$^{53}$
J. Vink\'{o},$^{34,40,46,17}$
X. Wang,$^{44,48}$
K. Zhang,$^{49}$\and
G. Zsidi,$^{34,46,54}$
\\
Affiliations are given at Section~\ref{sec:affiliation}
}
\date{Accepted 2021. Received 2021; in original form 2021}
\pubyear{2021}

\begin{document}
\label{firstpage}
\pagerange{\pageref{firstpage}--\pageref{lastpage}}
\maketitle
\begin{abstract}
    SN 2017jgh is a type IIb supernova discovered by Pan-STARRS during the C16/C17 campaigns of the~\kep{} mission. Here we present the~\kep{} and ground based observations of SN 2017jgh, which captured the shock cooling of the progenitor shock breakout with an unprecedented cadence. This event presents a unique opportunity to investigate the progenitors of stripped envelope supernovae. By fitting analytical models to the SN 2017jgh lightcurve, we find that the progenitor of SN 2017jgh was likely a yellow supergiant with an envelope radius of $\sim50-290~R_{\odot}$, and an envelope mass of $\sim0-1.7~M_{\odot}$. SN 2017jgh likely had a shock velocity of $\sim7500-10300$ km s$^{-1}$. Additionally, we use the lightcurve of SN 2017jgh to investigate how early observations of the rise contribute to constraints on progenitor models. Fitting just the ground based observations, we find an envelope radius of $\sim50-330~R_{\odot}$, an envelope mass of $\sim0.3-1.7~M_{\odot}$ and a shock velocity of $\sim9,000-15,000$ km s$^{-1}$. Without the rise, the explosion time can not be well constrained which leads to a systematic offset in the velocity parameter and larger uncertainties in the mass and radius. Therefore, it is likely that progenitor property estimates through these models may have larger systematic uncertainties than previously calculated.
\end{abstract}


\section{Introduction}\label{sec:introduction}
Massive stars end their lives as a core-collapse supernovae (CCSNe), leaving behind a black hole or neutron star and a supernova remnant. CCSNe come in a variety of observed sub-types depending on the properties of the progenitor star prior to explosion, the explosion mechanism itself, and its circumstellar environment. Type II supernovae display strong hydrogen lines in their early spectra, with this population consisting of Type II-P and II-L SNe (plateauing and linear lightcurve decay, respectively), Type IIn SNe (narrow line spectra), and Type IIb SNe~\citep[He in spectrum with disappearing H;][]{Filippenko1997}. Type I supernovae lack strong H lines, such as Type Ib SNe (He in spectrum), and Type Ic SNe~\citep[no He in spectrum;][]{Janka2012,GalYam2017}. SNe IIb along with Ib and Ic are known as stripped envelope supernova as they lose their hydrogen over time.

Weeks after the explosion of Type IIb Sne,  hydrogen lines begin to disappear and helium lines begin to dominate with the spectrum more closely resembling a type Ib supernova~\citep{Filippenko1997}, suggesting a progenitor which is mostly stripped of its hydrogen envelope. The mechanism which strips the hydrogen is still unknown, with possibilities including stellar winds, stellar rotation, binary interaction, and nuclear burning instabilities \citep[e.g.][]{Podsiadlowski1992,WoosleyLanger1993,BerstenBenvenuto2012,LiuModjaz2016,YoonDessart2017,BerstenFolatelli2018,NaimanSabach2019,SravanMarchant2019}. 

Type IIb SNe are of particular interest as many IIb SNe progenitors have been identified. Several have evidence of binary interaction (e.g. SN 1993J;~\citealt{MaundSmartt2004}, SN 2011dh;~\citealt{BerstenBenvenuto2012}). A well studied example of a type IIb supernova is SN 1993J~\citep{RichmondTreffers1994,NomotoSuzuki1993}. The progenitor of this supernova was identified from direct imaging as a late G or early K supergiant with an effective temperature between 4000-4500 K and an initial mass prior to explosion of 17 M$_{\odot}$. There is evidence that the progenitor of SN 1993J had a binary companion which may be responsible for stripping the progenitor of its hydrogen~\citep{MaundSmartt2004}.

Stripped supernovae can show two prominent peaks in their optical lightcurve. The first peak is a burst of emission after the explosion known as the shock cooling lightcurve~\citep[\scc{};][]{GalYam2017,Arcavi2017,SravanMarchant2020}, and the second peak is nuclear powered emission fueled by the decay of $^{56}$Ni.

The~\scc{} provides an opportunity to probe physical properties of the progenitor. This emission is produced when photons trapped behind optically dense material within the progenitor finally escape~\citep{WaxmanKatz2017}. The associated optical emission lasts on the order of days, and the shape of this curve depends on both the behaviour of the shock wave and the physical properties of the progenitor star (\citealt{NakarSari2010,WaxmanKatz2017} both provide a more complete review of shock breakout physics). As such, investigating the~\scc{} provides a unique opportunity to probe both the properties of the shock wave and the progenitor star.

There have been many supernovae observed with a~\scc{} - SN 1993J~\citep{RichmondTreffers1994}, SN 2011dh~\citep{ArcaviGalYam2011}, SN 2011fu~\citep{KumarPandey2013}, SN 2011hs~\citep{BufanoPignata2014}, SN 2013df~\citep{MoralesGaroffoloEliasRosa2014}, and SN 2016gkg~\citep{ArcaviHosseinzadeh2017,KilpatrickFoley2017}. Hydrodynamical modelling of SN 2011dh suggests that a star with a compact core and low mass envelope is most likely responsible for the observed~\scc~\citep{BerstenBenvenuto2012}. This was later confirmed by~\citet{NakarPiro2014} who employed semi-analytical models to show that a progenitor with an extended envelope was required. The~\citet{NakarPiro2014} model includes two relationships: one between the bolometric luminosity at the peak of the~\scc{} and the radius of the extended envelope, and the other between the time of the peak of the~\scc{} and the mass concentrated at the envelope's radius.~\citet{Piro2015} (hereafter P15) produced an analytical expression for the complete~\scc{}. P15 enforced conditions necessary to produce a~\scc{} which require that the material is massive enough for the shock to propagate and extended enough for the peak to be bright in the optical band. No explicit assumptions are made as to the density profile of the supernova progenitor. An alternative to the P15 model,~\citet{RabinakWaxman2011}, explicitly assume a polytropic density profile to provide a more physically realistic model of the progenitor. This was expanded upon by~\citet{SapirWaxman2017} (hereafter SW17) by scaling their model to better agree with hydrodynamical simulations, allowing the model to extend to later times.~\citet{PiroHaynie2020} (hereafter P20), like the P15 model, makes no assumptions about the density profile of the progenitor star, instead the P20 model makes use of a two-component velocity profile to attain better fits to the \scc{} than P15. All attempts at modelling the~\scc{} thus far rely on the assumption of hydrostatic equilibrium within the progenitor~\citep{Chevalier1992,NakarSari2010}.

Capturing the shock cooling lightcurve of a supernova is difficult owing to the short lifetime of these events. Most examples of~\scc{}s have only included the decline of the~\scc{}, sometimes managing to capture the peak as well. The rise of the~\scc{} was yet to be captured at high cadence, which presents a problem for the development and improvement of~\scc{} models and, thus, our ability to probe the progenitor of these supernovae. As mentioned in P15, the shape of the rise is highly dependant on the density structure of the progenitor which makes the rise difficult to accurately model. Without data of the~\scc{} rise, one cannot test how effective models are at this critical early stage~\citep{Piro2015,SapirWaxman2017,PiroHaynie2020}.

In order to effectively capture the complete~\scc{}, continuous, high cadence observations are needed. A cadence of < 1 day allows one to capture the earliest emission from the supernova, and high cadence observations are required in order to observe the rapidly evolving~\scc{}. Telescopes like the~\textit{Kepler Space Telescope} \citep[\kep;][]{HowellSobeck2014} and the~\textit{Transiting Exoplanet Survey Satellite} \citep[TESS;][]{RickerWinn2014} allow for this type of observing strategy. As such, we can expect more shock cooling lightcurves with higher cadence data to be observed, presenting an exciting opportunity to gain a much better understanding of the progenitors of type IIb supernovae~\citep{VallelyKochanek2021,FausnaughVallely2021}. In order to effectively use this new data, we need to determine which class of analytical~\scc{} models fits the data the best and how much of the~\scc{} must be observed to constrain progenitor models.

In this paper we present SN 2017jgh, a type IIb supernova which was observed by~\kep{} in Campaign 16.~\kep{} was able to capture the full lightcurve of SN 2017jgh, including the complete evolution of the~\scc{}. This is the first time a high cadence, complete shock cooling lightcurve has been observed. This represents a unique opportunity to evaluate the effectiveness of the analytical models and to investigate how important the rise is for accurate modelling of progenitor properties.
\section{Observations\label{sec:data}}
SN 2017jgh\footnote{EPIC: 211427218} was discovered by Pan-STARRS1~\citep[PS1;][]{ChambersMagnier2016} on 2017 Dec 23 at 11:09:36 (MJD 58110.965) with $g=20.21$ mag~\citep{ChambersHuber2017}. SN 2017jgh occurred at $(\alpha,\delta)=(09^{\text{h}}02^{\text{m}}56^{\text{s}}.736,+12\degree{}03'04''27)$, at a separation of 0.157 arcseconds from the centre of its host galaxy, 2MASX J09025612+1202596 with redshift 0.079. We estimated the Milky Way reddening towards SN 2017jgh as $E(B-V)=0.02$ mag~\citep{SchlegelFinkbeiner1998}. We do not take into account host galaxy extinction. This extinction is likely to redden the lightcurve, however the shape of the~\scc{} is more important than the colour when fitting so we believe the effects of this extinction to be negligible. We measure the radioactive maximum in the~\kep{} band to be $t_{max}=58127\pm1$ MJD. Throughout the paper, epochs are presented relative to this maximum, as $t-t_{max}$.
\begin{figure*}
    \includegraphics[width=\linewidth]{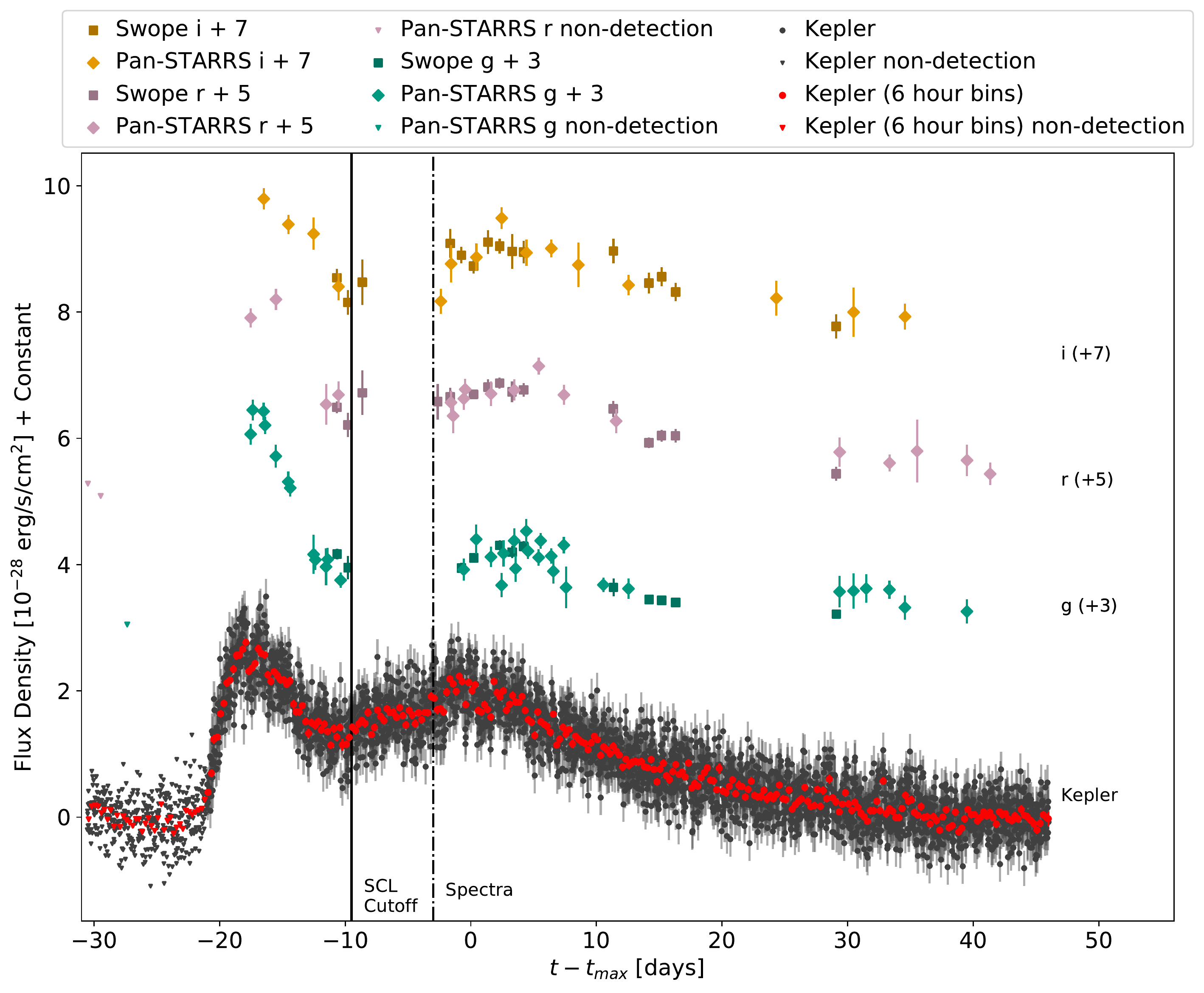}
    \caption{The combined~\kep{} and ground-based lightcurve of SN 2017jgh. The ground-based data is composed of Swope~\protect{\citep[Squares;][]{FolatelliPhillips2010}} and Pan-STARRS~\protect{\citep[Diamonds;][]{ChambersMagnier2016}} data. The black circles are the 30 minute cadence~\kep{} data. In order to better see the~\kep{} lightcurve, we include 6hr bins of this lightcurve (red circles). The solid vertical line indicates where the~\scc{} ends and the nuclear powered portion begins. The dashed vertical line shows when the spectrum of SN 2017jgh was taken.
    \label{fig:2017jghlightcurve}}
\end{figure*}
\begin{figure}
    \centering
    \includegraphics[width=\linewidth]{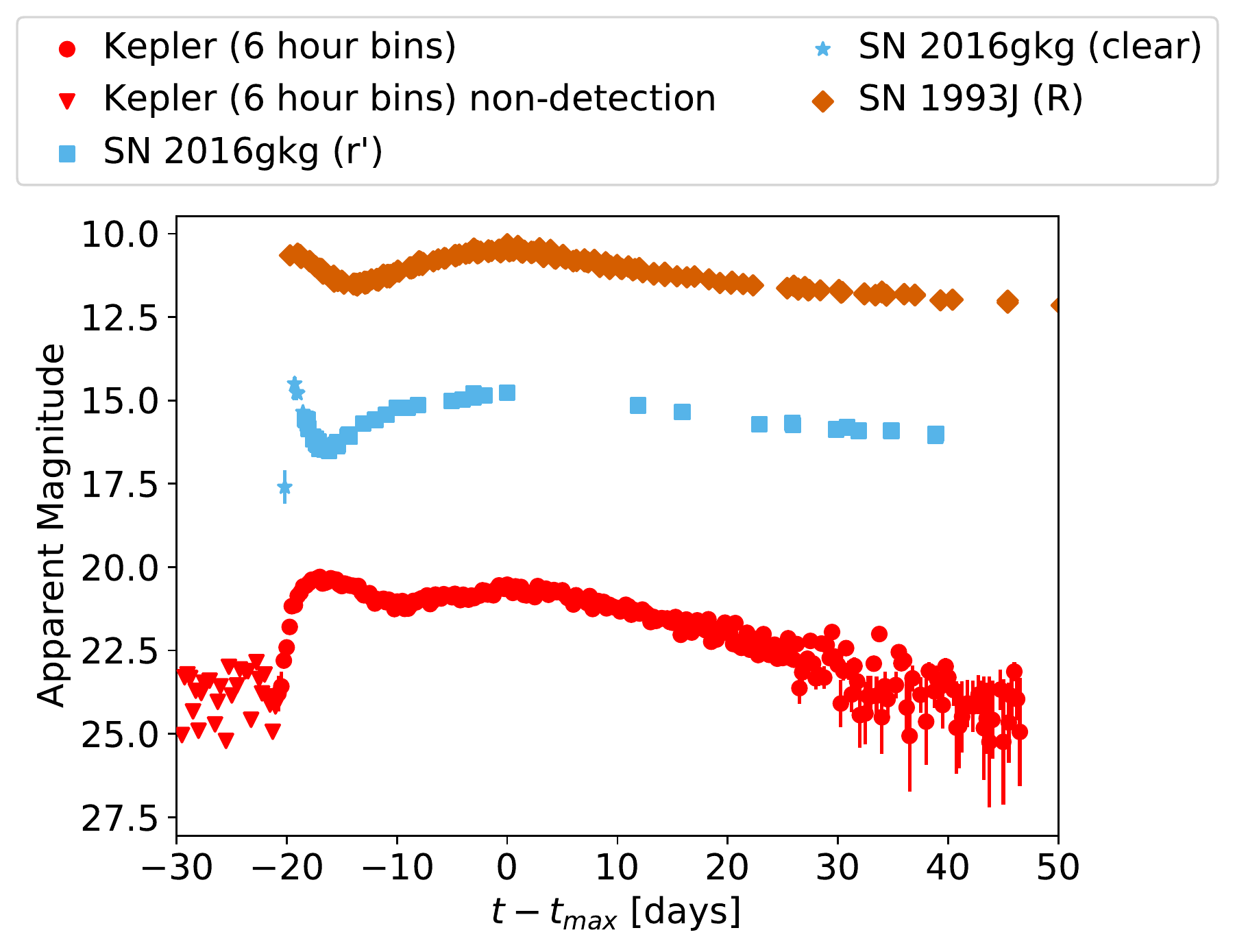}
    \caption{A comparison between the~\kep{} lightcurve of SN 2017jgh, the r'-band photometry of SN 2016gkg~\protect{\citep{ArcaviHosseinzadeh2017}}, and the R-band photometry of SN 1993J~\protect{\citep{RichmondTreffers1994}}. Clear filter observations are also included for SN 2016gkg~\protect{\citep{OteroBuso2016}}. The~\scc{} does not have the sharp rise and decline of SN 2016gkg and appears similar to SN 1993J.}
    \label{fig:comparison}
\end{figure}
\subsection{Ground-Based photometry}
Ground-based photometry was obtained from both PS1~\citep{ChambersMagnier2016,Dotson18} and the Swope Supernova Survey~\citep[SSS;][]{FolatelliPhillips2010}.

PS1 is a 1.8m telescope located at Haleakala on Maui, Hawaii. The telescope contains a 1.4 gigapixel camera, GPC1, mounted at the Cassegrain~\textit{f}/4.4 focus. GPC1 has sixty orthogonal transfer array devices, each with a 4846$\times{}$4846 pixel detector area. Each pixels measures 10 $\mu$m in size, giving a focal plane of 418.88 mm in diameter (3.3\textdegree{}). This gives a field of view 7.06 square degrees. PS1's filter system ($grizy$) is similar to SDSS~\citep{AbazajianAdelmanMcCarthy2009}, with the addition of a composite $gri$ 'wide' filter $w$~\citep{ChambersMagnier2016,TonryStubbs2012}.

Reduction of PS1 images (described in detail in~\citet{MagnierChambers2020}) is performed by the PS1 Image Processing Pipeline~\citep[IPP;][]{AMOS2006,MagnierLiu2008} which includes standard reductions, astrometric solution, stacking of nightly images, source detection, and photometry. The stacks are then sent to the Harvard FAS Cannon Research Computing cluster where the {\tt photpipe} pipeline~\citep{RestStubbs2005} performs difference imaging and transient identification.

SSS's optical photometry of SN 2017jgh  was obtained in $gri$ with the Swope 1-meter telescope at Las Campanas, Chile, from 17 Dec 2017 to Feb 7 2018. Following the description in~\citet{KilpatrickFoley2018}, all image processing and optical photometry on the Swope data was performed using {\tt photpipe}~\citep{RestStubbs2005}. The photometry were calibrated using standard sources from the Pan-STARRS DR1 catalog~\citep{FlewellingMagnier2020} in the same field as SN 2017jgh and transformed following the Supercal method~\citep{ScolnicCastertano2015} into the Swope natural system~\citep{KrisciunasContreras2017}. Deep $gri$ template images of the SN 2017jgh field were obtained on Jan 8 2019 and Jan 14 2019. Final image subtraction was performed using {\tt hotpants}~\citep{Becker2015}. Forced photometry was performed on the subtracted images.

\subsection{\kep{} Photometry}
 The K2 mission consists of a series of campaigns which observe different fields within the ecliptic plane.~\kep{} uses a 0.95 m aperture Schmidt telescope, orbiting the Earth in a heliocentric orbit. The telescope has one broad filter spanning 4000 to 9000\AA{}, with peak transmission roughly coinciding with the Pan-STARRS $r$-band~\citep{BrysonTenenbaum2010}.

The primary advantage of~\kep{} is the 30 minute cadence. This rapid cadence, combined with the roughly 80 day campaigns, allows for early and detailed lightcurves of transient events. 
Throughout campaign 16, the~\textit{Kepler} spacecraft had been running with the loss of 2 reaction wheels, so was operating on 2 remaining wheels and thrusters. This introduced a number of systematic effects into the lightcurve. These included reaction wheel jitter, which introduces additional short term noise, and solar pressure induced drifting, which introduced long term systematics with a characteristic "sawtooth" pattern.

As described in~\citet{ShayaOlling2015}, the data reduction pipeline begins  by correcting the CCD images for bias level, dark current, smear, nonlinear gain, undershooting pixels, and flat fields~\citep{QuintanaJenkins2010}. After initial calibration, the Presearch Data Conditioning (PDC)~\citep{StumpeSmith2012} applies corrections for both instrumental and spacecraft anomalies, and removes contamination from nearby stars.  During this process, a set of 14 cotrending basis vectors (CBVs) are generated by singular value decomposition to represent correlated instrumental artifacts such as flexing of the telescope structure, thermal transients, and drifting which occurs due to solar pressure. The PDC lightcurve then has a superposition of these CBVs removed such that the root-mean-squared (rms) deviations are minimised.  However, for SNe, this process would remove most of the physical variations in the light curve.  We therefore carefully create a set of CBVs using only quiet galaxies that are on the same CCD channel as the SN itself.   Then we solve for the coefficients of a superposition of CBVs that minimise the rms deviations before the onset of the SN event.

In order to calculate the~\kep{} zero point we follow the~\textit{Kepler Instrument Handbook}~\citep{KeplerInstrumentHandbook} which gives the following equation for the~\kep{} magnitude ($K_p$) in terms of the $(g-r)$ colour (where $g$ and $r$ are the SDSS filters):
\begin{gather}
    K_{p} = \begin{cases}
    r + 0.2(g-r) &\text{if $(g-r)\le0.8$ mag} \\
    r + 0.1(g-r) &\text{if $(g-r) > 0.8$ mag}
    \end{cases}
\end{gather}
The $(g-r)$ colour is calculated from the ground-based photometry. Once the psuedo-\kep{} magnitude is calculated, it is compared to the raw~\kep{} flux in order to calculate the~\kep{} zeropoint. We calculate the~\kep{} zeropoint to be $25.3\pm0.1$. We validate this zeropoint by calculating synthetic Kepler photometry using the Kepler bandpass listed on SVO filter profile service~\citep{RodrigoSolano2012,RodrigoSolano2020}, normalised to AB magnitudes. Since the GMO-S spectrum discussed in Section \ref{sec:spec} did not cover the full wavelength range, we use the SN 2016gkg spectra taken 2 days before radioactive maximum by \citep{JhaVanWyk2016}, normalised to the ground-based photometry for SN 2017jgh at the same phase. Comparing to the Kepler counts, we find a zeropoint of $23.33\pm0.15$, which is within 10\% of the zeropoint derived from ground-based photometry. Since there are unknown differences between the SN 2017jgh and SN 2016gkg spectra we use the ground-based photometric zeropoint of $25.3\pm0.1$.

\begin{figure}
    \centering
    \includegraphics[width=\linewidth]{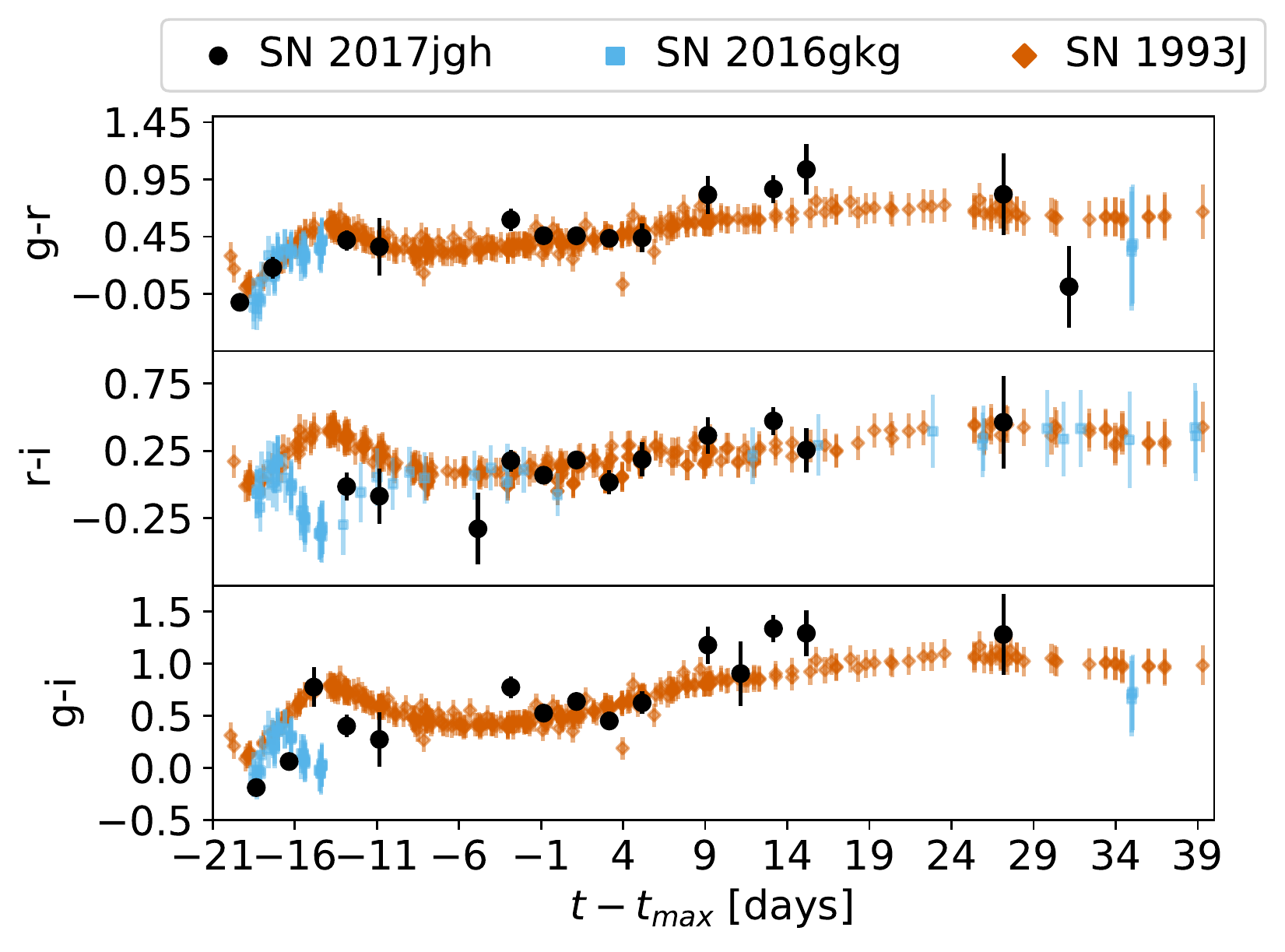}
    \caption{Colour evolution of SN 2017jgh (black), SN 2016gkg (blue), and SN 1993J (orange). The colour evolution of SN 2017jgh appears most similar with the SN 1993J evolution.}
    \label{fig:ColourEvolution}
\end{figure}

The combined~\kep{} and ground-based lightcurve is presented in Figure~\ref{fig:2017jghlightcurve}. The double-peaked profile is evident in the~\kep{} photometry and is mirrored in the ground-based data, although the ground-based data is significantly sparser and does not cover the rise. A comparison between SN 2017jgh, SN 1993J~\citep{RichmondTreffers1994} and SN 2016gkg~\citep{ArcaviHosseinzadeh2017} is presented in Figure~\ref{fig:comparison}, and the evolving colours of SN 2017jgh, SN 2016gkg and SN 1993J are shown in Figure~\ref{fig:ColourEvolution}\footnote{\label{note:snespace}The lightcurves and spectra of SN 2016gkg and SN 1993J were gathered from \url{https://sne.space/}~\protect{\citep{GuillochonParrent2017}}}. 

The lightcurve of SN 2017jgh appears most similar to SN 1993J, which has a very similar~\scc{} decline. The~\scc{} of SN 2016gkg has a much sharper decline, which may indicate it had less extended material. Overall, this suggests that the progenitor of SN 2017jgh was closer to the yellow supergiant progenitor of SN 1993J than the blue supergiant progenitor of SN 2016gkg. Additionally SN 1993J's colour evolution closely matches the g-r colour evolution of SN 2017jgh, which again suggests that the progenitor of SN 1993J and SN 2017jgh were very similar.

An excerpt of the lightcurve data is shown in Table~\ref{tab:lightcurve_data}, with the full dataset available online.
\begin{table*}
    \begin{center}
    \caption{Excerpt of lightcurve data for SN 2017jgh. The full lightcurve is available as supplementary material.}
    \label{tab:lightcurve_data}
    \begin{tabular}{ccccc}
        Time (MJD)  &   Flux (erg s$^-1$ cm$^-2$    &    Flux Error (erg s$^-1$ cm$^-2$)    & Band  & Instrument\\
        58117.33 &   1.50e-28      &    9.16e-30              & r     & Swope\\
        58117.34 &   1.54e-28      &    1.41e-29               & i     & Swope\\
        58117.35 &   1.17e-28       &    8.38e-30              & g     & Swope\\
        58118.21 &   1.21e-28      &    1.89e-29             & r     & Swope\\
        58118.22 &   1.15e-28      &    1.95e-29             & i     & Swope\\
        58118.22 &   9.51e-29       &    1.85e-29               & g     & Swope\\
        58119.36 &   1.72e-28      &    3.54e-29             & r     & Swope\\
        58119.36 &   1.47e-28      &    3.61e-29             & i     & Swope\\
        58125.37 &   1.58e-28      &    2.82e-29              & r     & Swope\\
        58126.36 &   1.66e-28      &    1.46e-29             & r     & Swope\\
    \end{tabular}
    \end{center}
\end{table*}

\subsection{Spectroscopy} \label{sec:spec}
\begin{figure}
    \includegraphics[width=\linewidth]{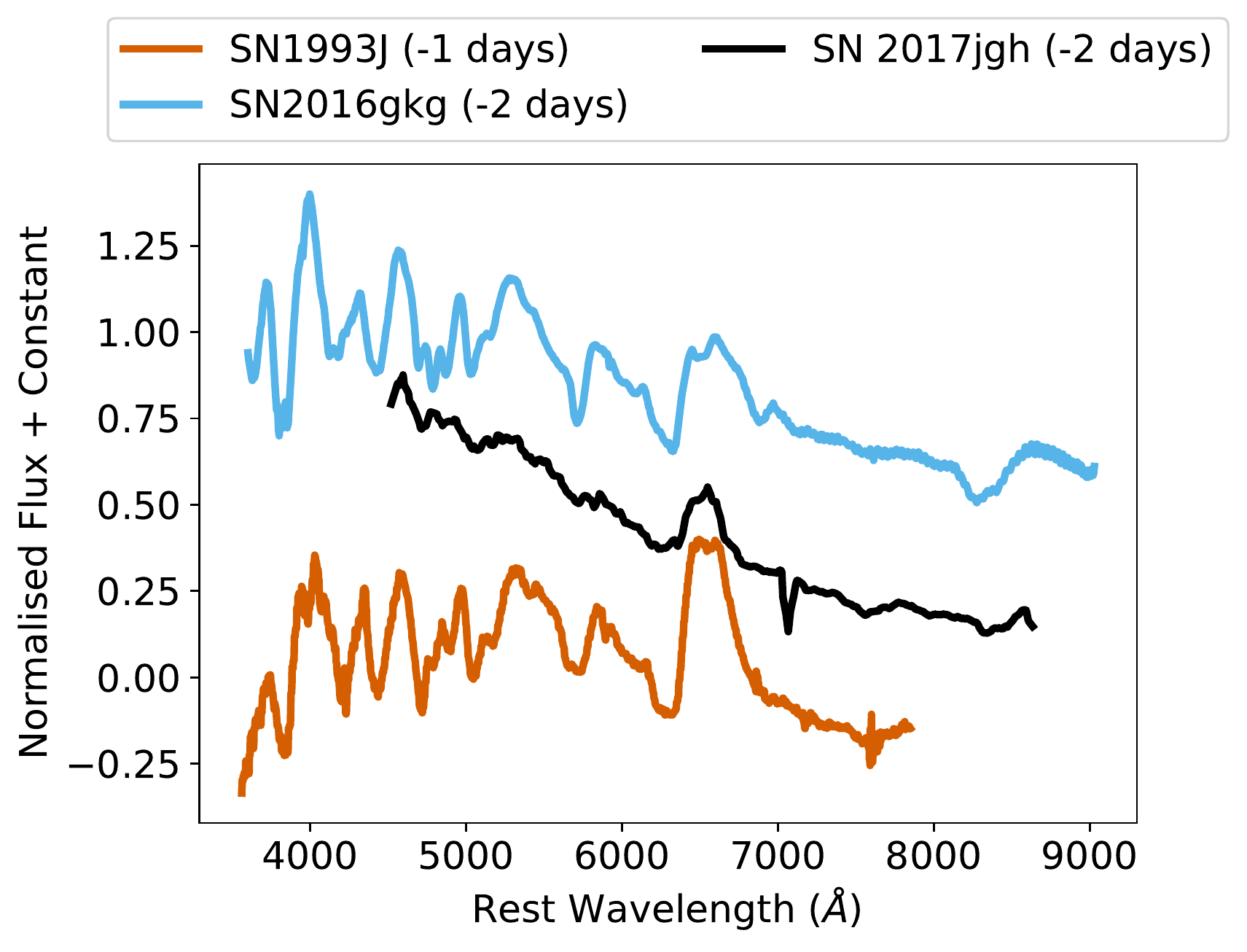}
    \caption{The spectrum of SN 2017jgh 2 days before radioactive maximum, compared to SN 1993J 1 day before radioactive maximum~\protect{\citep{JefferyKirshner1994}} and SN 2016gkg 2~days before radioactive maximum~\protect{\citep{JhaVanWyk2016}}. All of the spectra have been normalised and then shifted in order to easily differentiate between them. }
    \label{fig:SN 2017jghSpectra}
\end{figure}
We obtained optical spectroscopy on Jan 7th 2018, two days before radioactive maximum (14 day after discovery) using the Gemini Multi-Object Spectrograph on the Gemini South telescope \citep[GMOS-S;][]{HookJorgensen2004}. GMOS-S was configured using the R400 grating and a 1 arcsec longslit, with the detectors in 2x2 binning mode, covering a spectral range from 475 to 925 nm at a resolution of~$\sim{}$900. Data reduction was completed using the Gemini IRAF package. The data was first corrected for bias and flat-fielded using standard techniques. Cosmic rays were removed~\citep{vanDokkum2001}, as well as any detector bad columns.  Note that the width of a cosmic ray detection or bad column is small compared to the size of the resolved features in our spectra. The sky background was removed by first subtracting nodded pairs of images, followed by removing any residual by linear fitting the sky background near the target. One-dimensional spectra were then produced using variance weighted extraction, utilizing a 1 arcsec spatial aperture. A similarly processed standard star, EG-131, was used to correct each extracted spectra for instrument response, prior to producing the final co-added spectrum. This Gemini baseline correction does not provide an absolute flux-calibrated spectrum, as the standard is not contemporaneous with the science observation, nor are slit-losses or second-order contamination a part of the standard correction. These calibration issues will likely compound at shorter wavelengths, which may be responsible for the increased flux around 5000~\AA in the SN 2017jgh spectrum compared to similar SNe.

The reduced Gemini spectrum is shown in Figure~\ref{fig:SN 2017jghSpectra}. The spectrum of SN 2017jgh is compared to both SN 1993J at 1 day before radioactive maximum (\citealp{JefferyKirshner1994}; note that they state the spectra is `not photometrically accurate'), and SN 2016gkg at 2 days before radioactive maximum~\citep{JhaVanWyk2016}. The red ($>6000$\AA{}) spectrum of SN 2017jgh is similar to both the SN 1993J spectrum, and the SN 2016gkg spectrum. Below 6000\AA{} the spectrum of SN 2017jgh diverges from the other supernovae, likely due to the calibration issues we have described earlier.

We used the Supernova Identification~\citep[SNID;][]{BlondinTonry2007} program which classified SN 2017jgh as a Type IIb supernova. However it is worth noting that with only one spectrum we do not have the robust IIb classification criteria of the observation of strong He features and the disappearance of H features. 

\section{Shock Cooling Curve Models}\label{sec:methods}
Here we summarise the formalism of the P15, P20, and SW17 analytical models.
\subsection{P15 Model}
The P15 model makes no assumptions about the density profile of the progenitor and assumes a simple expanding photosphere. P15 used these assumptions along with simple thermodynamic equations to derive the bolometric luminosity of a~\scc{}. We follow~\citet{ArcaviHosseinzadeh2017}, who recast original equation for luminosity in P15 into the following equation which depends on the radius of the extended envelope ($R_{e}$), the mass of the envelope ($M_{e}$), and the velocity of the envelope ($v_{e}$):
\begin{dmath}
\label{eq:PiroLuminosity}
\frac{L(t)}{\text{erg s$^{-1}$}} = \frac{8.27\times{}10^{42}}{\text{cm}}\kappa_{0.34}^{-1}v_{9}^{2}R_{13}\left(\frac{M_{c}}{M_{\odot}}\right)^{0.01}\times\\
\exp\left[\frac{-4.135\times{}10^{-11}}{\text{cm g$^{-1}$ s$^{-1}$}}t(tv_{9}+2\times{}10^{4}R_{13})\times\\
\kappa_{0.34}^{-1}\left(\frac{M_{c}}{M_{\odot}}\right)^{0.01}\left(\frac{M_{e}}{0.01M_{\odot}}\right)^{-1}\right]
\end{dmath}

\noindent
where $\kappa_{0.34}$ is the opacity in $0.34~\rm{}cm^{2}\,g^{-1}$, $v_{9}$ is the envelope velocity in $10^{9}\rm{}cm\,s^{-1}$, $R_{13}$ is the envelope radius in $10^{13}\rm{}cm$, and $t$ is the time since explosion in seconds. Both $M_{e}$ and $M_{c}$ (the core mass) are in solar masses.\lbreak{}
Assuming the emission is a blackbody with expanding radius $R(t)=R_{e}+v_{e}t$, we can estimate the temperature as:
\begin{dmath}
\label{eq:PiroTemp}
\frac{T(t)}{\text{K}} = \left[\frac{L(t)}{4\pi{}R^{2}(t)\sigma}\right]^{1/4}
\end{dmath}

\noindent
where $\sigma$ is the Stefan-Boltzmann constant. With both the bolometric luminosity and the temperature defined, we are able to calculate the observed flux in arbitrary bands.
\subsection{P20 Model}
P20 provide an improvement upon the P15 model by considering a two component model. This model consists of outer material with a steep velocity gradient, and inner material with a shallow velocity gradient. The transition between these layers occurs at the transition velocity $v_{t}$, related to the total energy of the extended material $E_{e}$ by:
\begin{gather}
\label{eq:V_to_E}
    v_{t} = \sqrt{\frac{(n-5)(5-\delta)}{(n-3)(3-\delta)}}\sqrt{\frac{2E_{e}}{M_{e}}}
\end{gather}
Where $t$ is the time in days, $n$ and $\delta$ are numerical factors which control the radial dependence of the outer and inner density structure, respectively. P20 show that the solution is not sensitive to the values of $n$ and $\delta$ as long as $n\gg1$ and $\delta\gtrsim1$. They suggest using typical value of $n\approx10$ and $\delta\approx1.1$. The P20 model has a two component luminosity -- the luminosity of the inner region and the luminosity of the outer region. Within the inner region, the luminosity is defined as:
\begin{gather}
    \frac{L(t\le{}t_{d})}{\text{erg s$^{-1}$}}\approx1.157\times10^{-5}\frac{E_{th}(v_{t}, t_{d})}{t_{d}}\left(\frac{t_{d}}{t}\right)^{4/(n-2)}
\end{gather}
where $t_{d}$ is the time when the~\scc{} reaches the transition between the inner and outer regions, and is equal to:
\begin{gather}
    \frac{t_{d}}{\text{days}} = 5.16\times10^{6}\sqrt{\frac{3\kappa{}KM_{e}}{(n-1)v_{t}c}}
\end{gather}
where $c$ is the speed of light in km s$^{-1}$, and the numerical factor $K$ is equal to $\frac{(n-3)(3-\delta)}{4\pi(n-\delta)}$; for our values of $n$ and $\delta$, $K=0.119$. $E_{th}(v_{t}, t_{d})$ is the thermal energy at a given velocity and time:
\begin{gather}
    \frac{E_{th}(v_{t}, t_{d})}{\text{erg}} = 5.147\times10^{30}t_{d}\frac{\pi(n-1)}{3(n-1)}\frac{cR_{e}v_{t}^{2}}{\kappa}
\end{gather}
The luminosity of the outer region is:
\begin{gather}
    \frac{L(t\ge{}t_{d})}{\text{erg s$^{-1}$}} = 1.157\times10^{-5}\frac{E_{th}(v_{t},t_{d})}{t_{d}}\exp\left[-\frac{1}{2}\left(\frac{t^{2}}{t_{d}^{2}}-1\right)\right]
\end{gather}
In order to calculate the luminosity in any band we once again assume the~\scc{} radiates as a black body, with a temperature:
\begin{gather}
    \frac{T}{\text{K}} = \left(\frac{L}{4\pi{}r_{ph}^{2}\sigma}\right)^{1/4}
\end{gather}
Here $r_{ph}$ is the photospheric radius (the point where the optical depth $\tau$ is equal to $2/3$). The photospheric radius is also a two-component function dependent on $t_{ph}$, the time when the photosphere reaches the transition between the inner and outer regions.
\begin{align}
    \frac{t_{ph}}{\text{days}} &= 5.16\times10^{6}\sqrt{\frac{3\kappa{K}M_{e}}{2(n-1)v_{t}^{2}}} = \sqrt{\frac{c}{2v_{t}}}t_{d} \\
    \frac{r_{ph}(t\le{}t_{ph})}{\text{cm}} &= 8.64\times10^{9}\left(\frac{t_{ph}}{t}\right)^{2/(n-1)}v_{t}t \\
    \frac{r_{ph}(t\ge{}t_{ph})}{\text{cm}} &= 8.64\times10^{9}\left[\frac{\delta-1}{n-1}\left(\frac{t^{2}}{t_{ph}^{2}}-1\right)+1\right]^{-1/(\delta-1)}v_{t}t
\end{align}
The free parameters of the P20 model are $M_{e}$ in solar masses, $R_{e}$ in solar radii and $v_{t}$ in km s$^{-1}$ (or $E_{e}$ as they are related by equation~\ref{eq:V_to_E}).
\subsection{SW17 Model}
\citet{RabinakWaxman2011} assume a polytropic progenitor density profile to derive an analytical form of the early lightcurve, characterised by the polytropic index $n$. This allows us to differentiate between progenitors with a convective envelope and a polytropic index of $n=3/2$, such as red supergiants (RSG), and progenitors with a radiative envelope and a polytropic index of $n=3$, such as blue supergiants (BSG).~\citet{SapirWaxman2017} improve upon the \citet{RabinakWaxman2011} model by introducing an additional term that suppresses the luminosity at later times. This accounts for the shock phase days after the initial explosion, when the shock cooling emission begins to emerge from the inner layers. The SW17 bolometric luminosity (after the recasting done by \citet{ArcaviHosseinzadeh2017}\footnote{Note that \citet{ArcaviHosseinzadeh2017} had a typo in their equation 6 as the factor of 19.5 was included in the square root term within the exponent. This has been corrected here.}) is:
\begin{align}
\label{eq:WaxmanLuminosity}
\frac{L_{\text{n=3/2}}(t)}{\text{erg s$^{-1}$}} &= \frac{1.88\times10^{42}}{\text{cm}}\times\\&\left(\frac{v_{s,8.5}t^{2}}{f_{p}M\kappa_{0.34}}\right)^{-0.086}\frac{v_{s,8.5}^{2}R_{13}}{\kappa_{0.34}}\times\nonumber\\&\exp{\left(-\left\{\frac{1.67t}{\frac{19.5}{\text{days cm$^{-0.5}$ s$^{-0.5}$}}(\kappa_{0.34}M_{e}v^{-1}_{s,8.5})^{0.5}}\right\}^{0.8}\right)\nonumber}\\
\frac{L_{\text{n=3}}(t)}{\text{erg s$^{-1}$}} &= \frac{1.66\times10^{42}}{\text{cm}}\times\\&\left(\frac{v_{s,8.5}t^{2}}{f_{p}M\kappa_{0.34}}\right)^{-0.175}\frac{v_{s,8.5}^{2}R_{13}}{\kappa_{0.34}}\times\nonumber\\&\exp{\left(-\left\{\frac{4.57t}{\frac{19.5}{\text{days cm$^{-0.5}$ s$^{-0.5}$}}(\kappa_{0.34}M_{e}v^{-1}_{s,8.5})^{0.5}}\right\}^{0.73}\right)\nonumber}
\end{align}
Here $v_{s,8.5}$ is the shock velocity in $10^{8.5}~\rm{}cm\,s^{-1}$, $M$ is equal to $M_{e}+M_{c}$, and $t$ is the time since explosion in days. The factor $f_{p}$ is equal to $\sqrt{\frac{M_{e}}{M_{c}}}$ for n=3/2, and $0.08(M_{e}/M_{c})$ for n=3.\lbreak{}
The temperature is given as:
\begin{align}
\label{eq:WaxmanTemp}
\frac{T_{n=3/2}(t)}{\text{K}} &= 2.05\times{}10^{4}\times\\
&\left(\frac{v^{2}_{s,8.5}t^{2}}{f_{p}M\kappa_{0.34}}\right)^{0.027}\left(\frac{R_{13}}{\kappa_{0.34}}\right)^{0.25}t^{-0.5}\nonumber\\
\frac{T_{n=3}(t)}{\text{K}} &= 1.96\times{}10^{4}\times\\
&\left(\frac{v^{2}_{s,8.5}t^{2}}{f_{p}M\kappa_{0.34}}\right)^{0.016}\left(\frac{R_{13}}{\kappa_{0.34}}\right)^{0.25}t^{-0.5}\nonumber
\end{align}
Under the assumption that the emission is characterised by a black body, this once again allows us to calculate the luminosity in any band.

We make a number of assumptions for all models, which follow the assumptions made by~\citet{ArcaviHosseinzadeh2017}. We assume $\kappa_{0.34}=1$, typical for solar composition materials to exhibit electron scattering. We also set $M_{c}=1M_{\odot}$. As stated in~\citet{ArcaviHosseinzadeh2017}, the early lightcurve is weakly dependant on this factor and our results are insensitive to it.

\section{Fitting the Shock Cooling Lightcurve}\label{sec:results}
For each model, we fit the $gri$ bands and the binned~\kep{} data simultaneously. We fit from the first observation up to -9.5 days before radioactive maximum. This was chosen to ensure our fits are not contaminated by emission from the main radioactive peak. For the P15, SW17, and P20 models, we fit the parameters $M_{e}$, $R_{e}$, $v_{e}$ or $v_{s}$ or $v_{t}$, respectively, and the offset time, $t_{\text{off}}$, between the earliest observation and the start of the~\scc{}.

We make use of the Python {\tt emcee} package~\citep{ForemanMackeyFarr2019}, which provides an implementation of an affine-invariant ensemble Monte Carlo Markov Chain (MCMC) sampler. This algorithm attempts to produce an approximation of the posterior given a model, data, and a likelihood function which states how well the model fits the data. A number of initial positions are randomly chosen and evaluated with the likelihood function. From these initial positions, walkers traverse the parameter space, at each step deciding to either move to a random new position or stay at their current position based on how well the new parameter position fits the data. After each step, the walkers record their position in a chain. After a large number of steps these chains will approximate the posterior. We use the reduced chi-squared as our likelihood function. We note that a chi-squared loss function naturally arises from a Gaussian log-likelihood, which assumed that each observations has white (uncorrelated) noise. The reduced chi-squared is inversely proportional to the degrees of freedom, which can be thought of as a regularization term to ensure that we are not biasing our fits to favor the higher cadence of the~\kep{} data.

For all fits, we use 500 walkers with a burn-in phase of 100 steps followed by 1000 additional steps. These were chosen after manually investigating the posterior and the walker chains to ensure they converged.
The uniform priors used are provided in Table~\ref{tab:mcmcparam}. These were initially chosen manually to fit the data while producing physically reasonable masses, radii, and velocities.
\begin{table}
\centering
    \caption{Uniform prior used for our MCMC fits.}
    \label{tab:mcmcparam}
        \begin{tabular}{@{}llll@{}}
            $R_{e}$ (R$_{\odot}$) & $M_{e}$ (M$_{\odot}$) & $v$ (km s$^{-1}$) & $t_{off}$ (Days) \\
            $0\to500$ & $0\to5$ & $0\to4\times10^{4}$ & $0\to15$
        \end{tabular}
\end{table}
\begin{figure}
    \includegraphics[width=\linewidth]{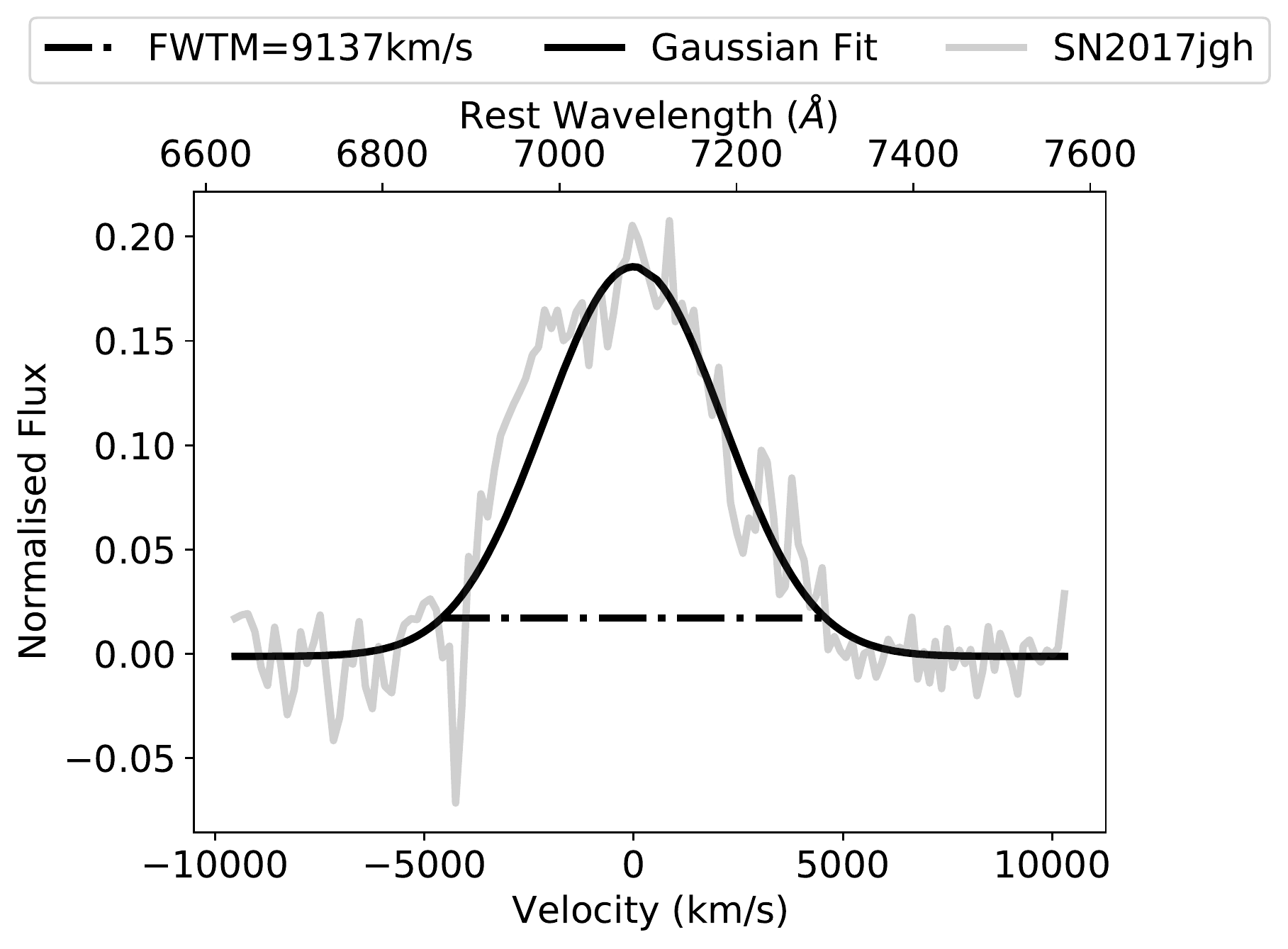}
    \caption{A Gaussian fit to the continuum subtracted H$\alpha$ emission of SN~2017jgh. The solid black line is our Gaussian fit and the dashed line the full width at tenth maximum, used to estimate the velocity.}
    \label{fig:spectrumfit}
\end{figure}
As an additional method of evaluating these fits, we compare their estimates of the velocity with that estimated from our spectrum. Following~\citet{JhaVanWyk2016}, we approximate the expansion velocity of the supernova from the full width at tenth max of the H$\alpha$ line. Figure~\ref{fig:spectrumfit} shows a Gaussian fit to the H$\alpha$ emission of SN 2017jgh in velocity space. The expansion velocity is measured to be $9100\pm470 km s^{-1}$.

\begin{figure*}
    \includegraphics[width=\linewidth]{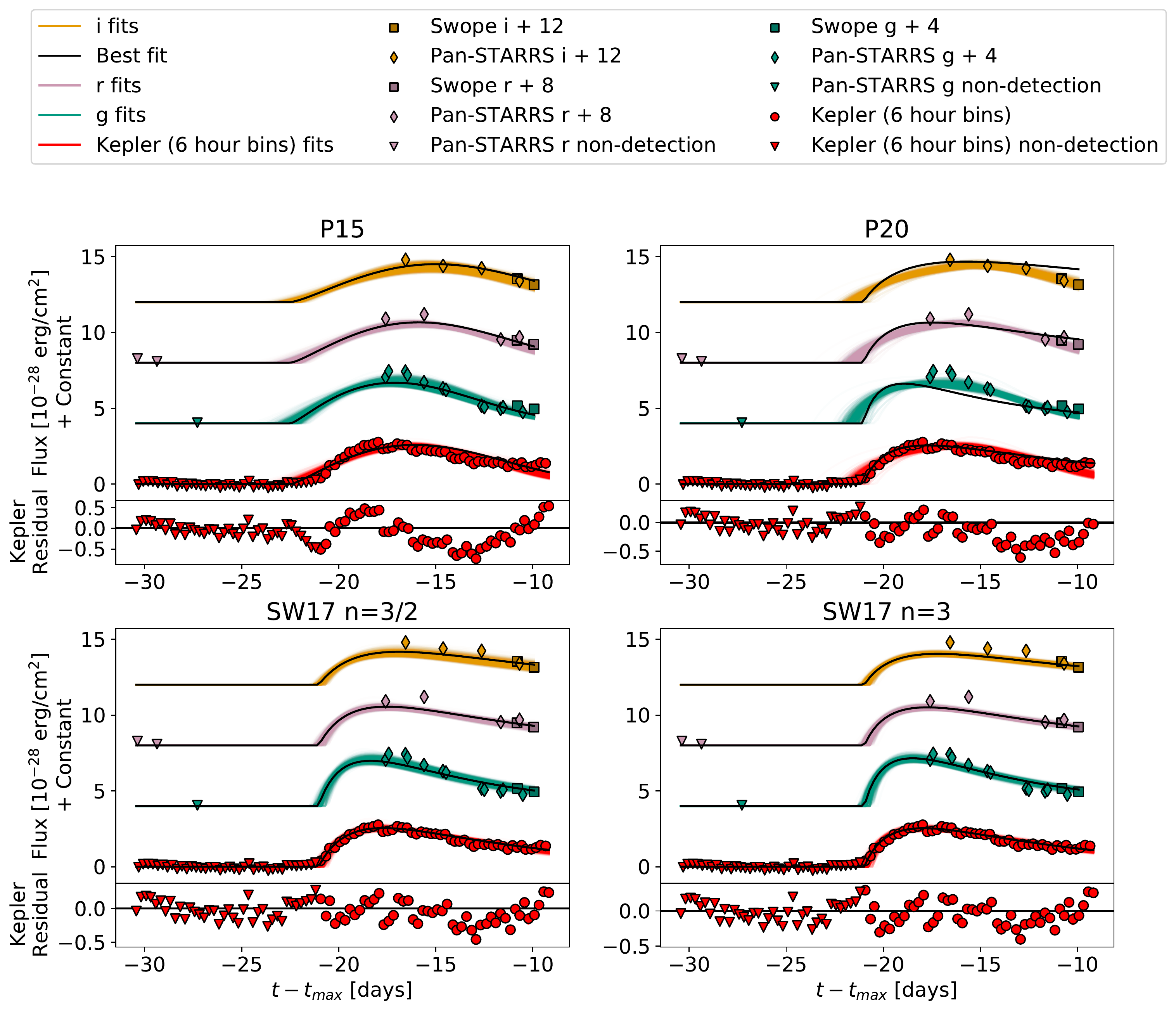}
    \caption{Model fits to the lightcurve of SN 2017jgh. The coloured lines are 1000 randomly drawn samples from the MCMC chain which give a visual understanding of the shape of the posterior. The black line is the median value of the posterior. Note that the median is taken from each parameter posterior independently so will differ from the randomly drawn samples. Additionally the P15 and P20 models contain non-Gaussian posteriors so the median model is not an accurate reflection of the best fitting model.}
    \label{fig:results}
\end{figure*}
\begin{figure*}
    \begin{subfigure}{0.49\textwidth}
    \includegraphics[width=\linewidth]{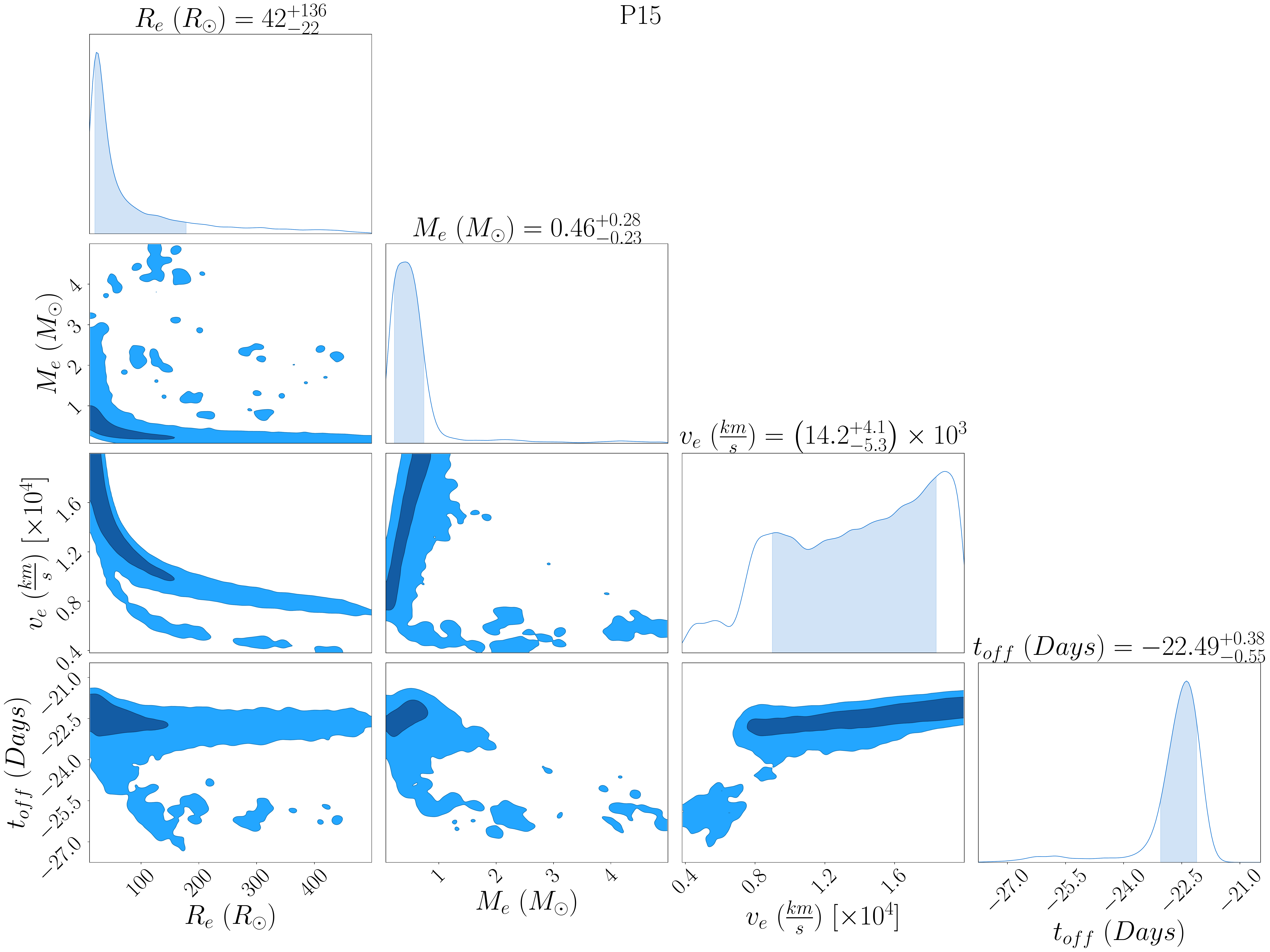}
    \end{subfigure}
    \begin{subfigure}{0.49\textwidth}
    \includegraphics[width=\linewidth]{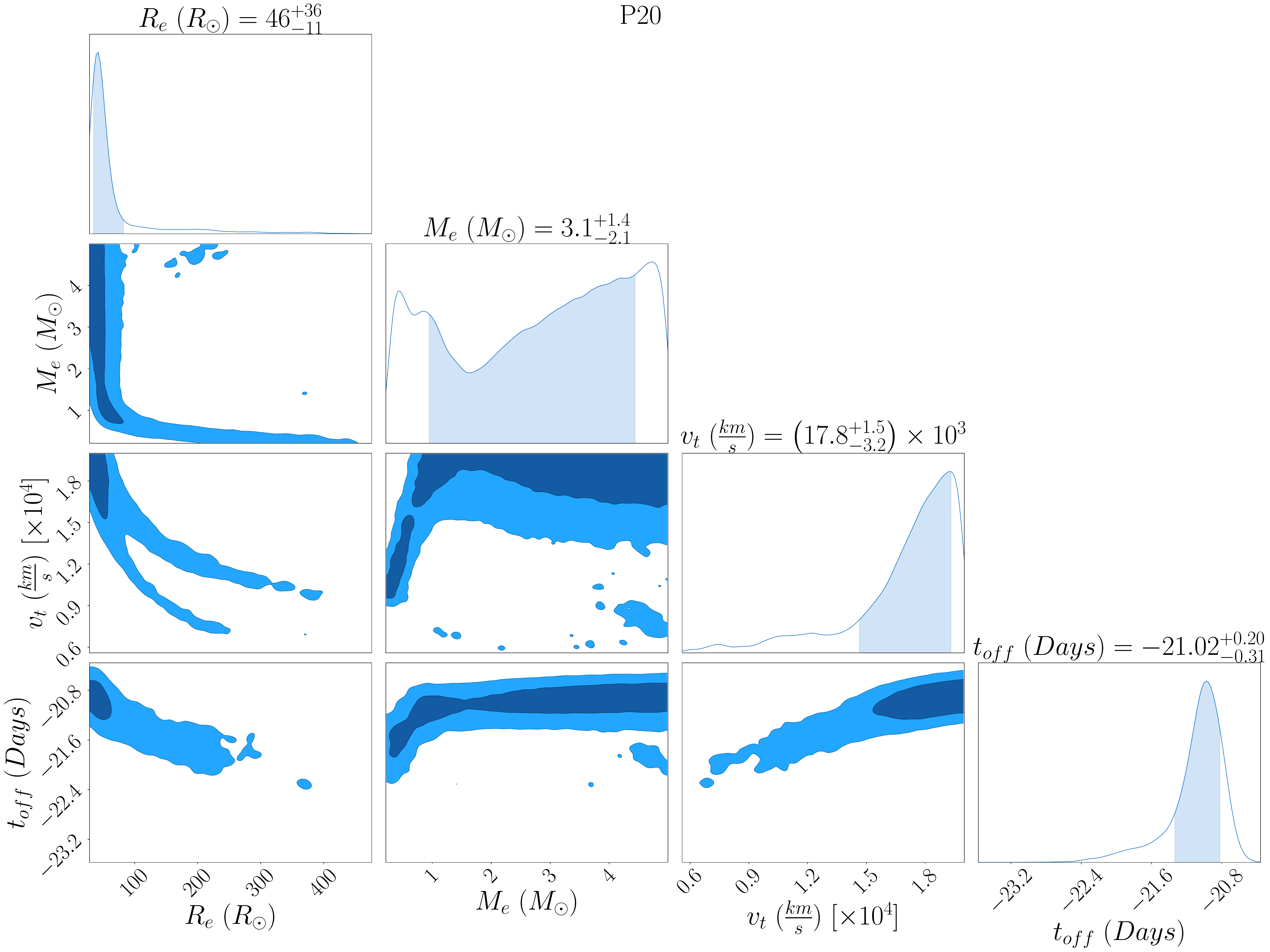}
    \end{subfigure}
    \begin{subfigure}{0.49\textwidth}
    \includegraphics[width=\linewidth]{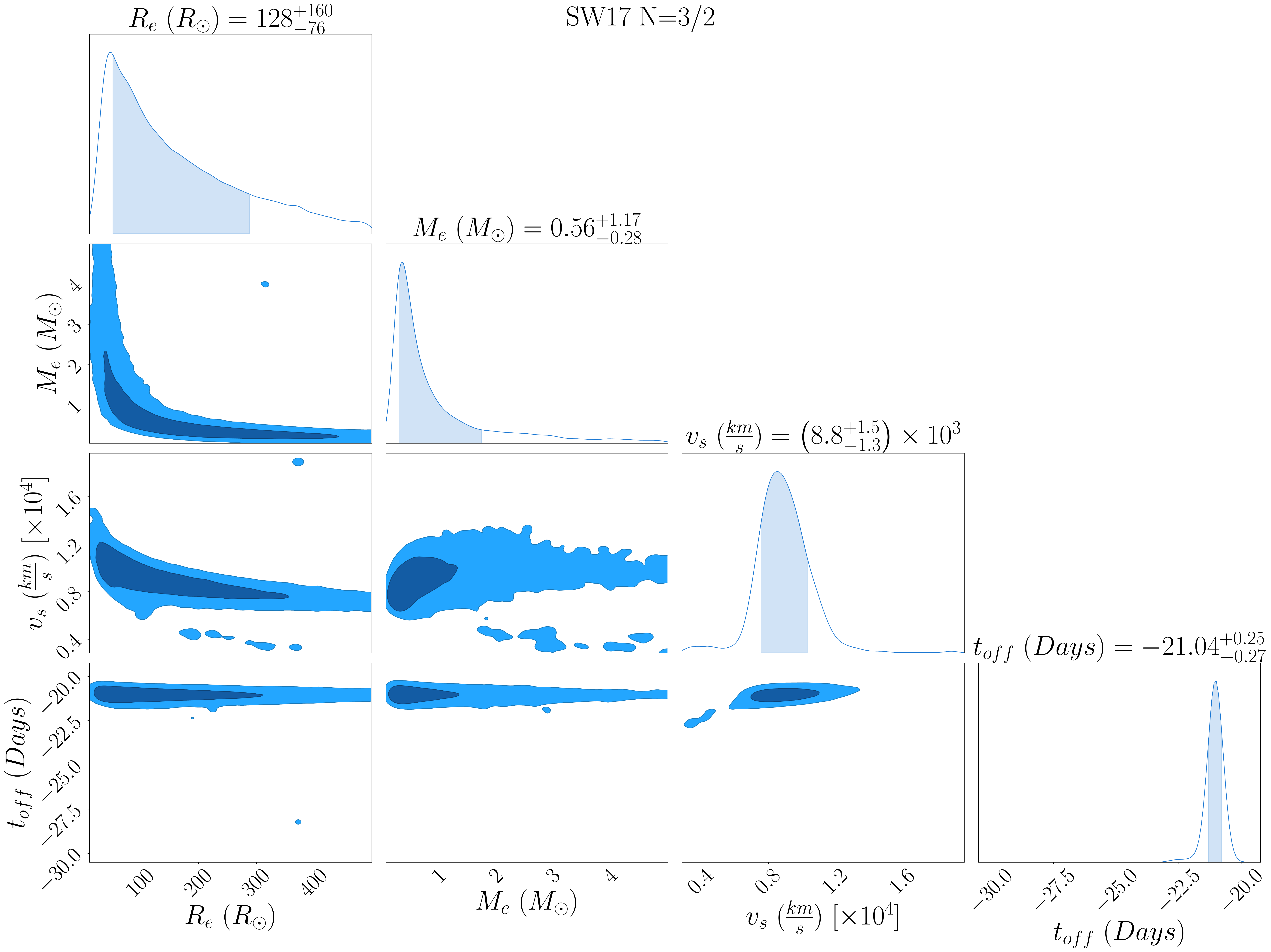}
    \end{subfigure}
    \begin{subfigure}{0.49\textwidth}
    \includegraphics[width=\linewidth]{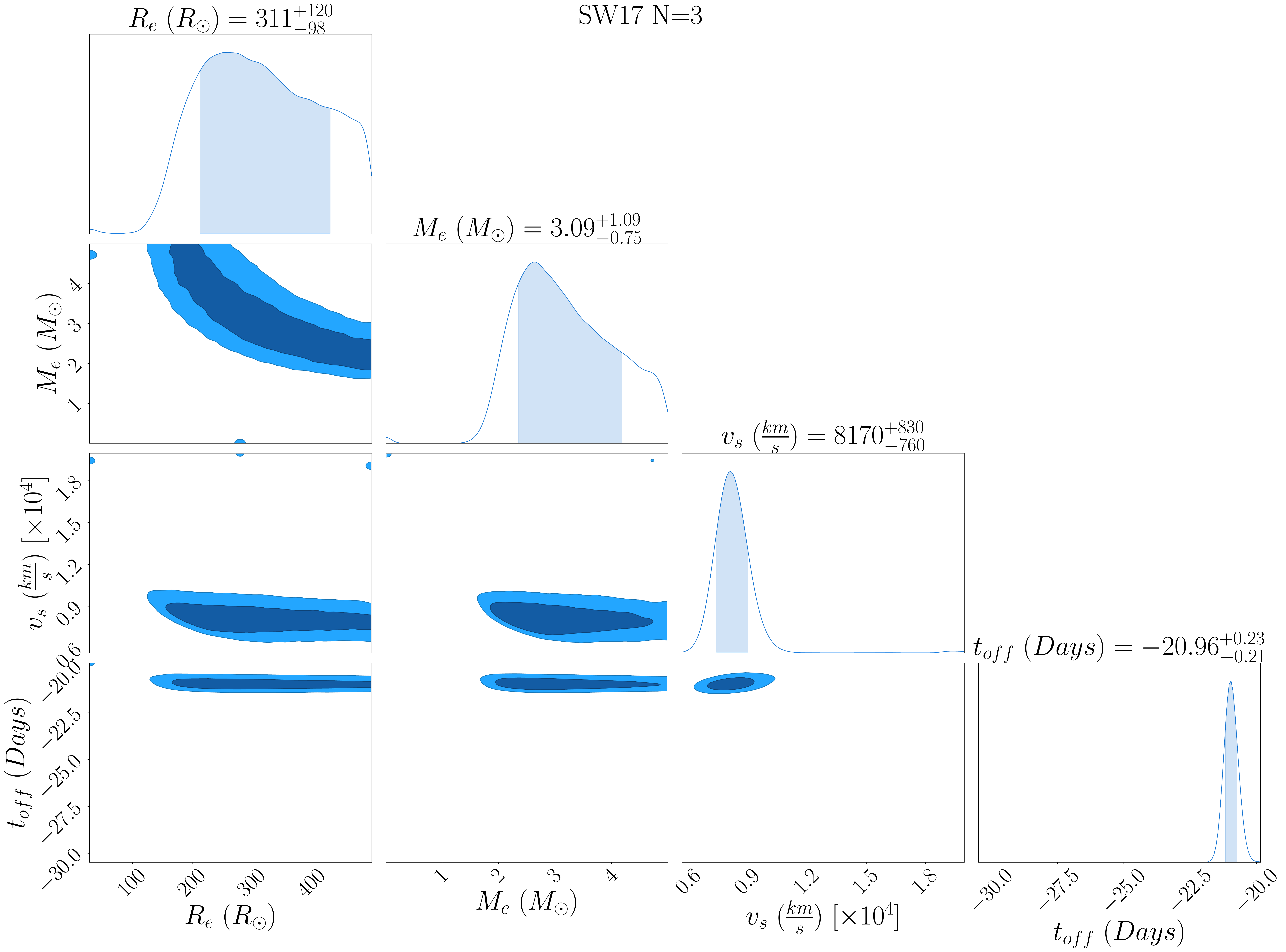}
    \end{subfigure}
    \caption{Corner plot \& results of our fits to the lightcurve of SN 2017jgh. The dark blue regions show the 1 sigma posterior. Created using the \protect{\href{https://samreay.github.io/ChainConsumer/index.html}{chainconsumer}} API. The best fitting value was calculated using the~\textit{cumulative} method of chainconsumer, which takes the 50th percentile of the parameter posterior as the best fitting value. The 16th and 84th percentiles are then the lower and upper bounds, respectively. The shaded region of each posterior, and the joint posteriors, show the lower and upper bounds (the 16th and 84th percentile). Note that $t_{off}$ is given in terms of days relative to radioactive maximum.}
    \label{fig:corner}
\end{figure*}
\begin{table*}
    \centering
    \begin{tabular}{c|cccccc}
         Model & $R_{e}~\left(R_{\odot}\right)$ & $M_{e}~\left(M_{\odot}\right)$ & $v~\left(10^{3}~\mathrm{km} \mathrm{s}^{-1}\right)$ & $t_{off}~\left(t-t_{max}\right)$ & Reduced $\chi^{2}$ & DOF \\
         P15 & $42.41^{+135.66}_{-22.32}$ & $0.46^{+0.28}_{-0.23}$ & $14.25^{+4.14}_{-4.26}$ & $-22.50^{+0.38}_{-0.55}$ & 17.79 & 4 \\
         P20 & $45.82^{+36.48}_{-11.04}$ & $3.07^{+1.37}_{-2.12}$ & $17.79^{+1.53}_{-3.16}$ & $-21.02^{+0.20}_{-0.31}$ & 12.82 & 4 \\
         SW17 n=3/2 & $127.82^{+160.35}_{-76.11}$ & $0.56^{+1.17}_{-0.56}$ & $8.81^{+1.50}_{-1.26}$ & $-21.05^{+0.25}_{-0.27}$ & 6.57 & 4 \\
         SW17 n=3 & $310.62^{+119.80}_{-97.59}$ & $3.09^{+1.09}_{-0.75}$ & $8.17^{+0.83}_{-0.76}$ & $-20.96^{+0.23}_{-0.21}$ & 6.39 & 4
    \end{tabular}
    \caption{Best fitting parameters for each model, found by taking the 16th, 50th, and 84th percentile of each parameter as the lower bound, best value, and upper bound, respectively. Note that this does not take into account the dependencies between parameters. The reduced chi squared of the best fitting value is also included. The SW17 models fit the lightcurve best, and have very similar velocity and offset times, however differ greatly in both the envelope radius and mass parameters. Since SW17 n=3/2 has a best fitting velocity which is less than one standard deviation of the spectral velocity (9100$\pm$470km s$^{-1}$), we choose this as the preferred model for SN 2017jgh.}
    \label{tab:results}
\end{table*}
The best fit to the lightcurve of SN 2017jgh is shown in Figure~\ref{fig:results} for each of our models of interest, the corner plots of each model are shown in Figure~\ref{fig:corner}, and the best fitting values are given in Table~\ref{tab:results}. The best fitting value is taken as the 50th percentile of the parameter posterior, with the 16th and 84th percentile as the lower and upper bounds, respectively. Note that this does not take into account the dependencies between parameters.

As can be seen in Figure~\ref{fig:results}, P15 appears to perform the worst of the four models, with a small divergence at the start of the~\scc{} rise and a larger deviation at about -15 days before radioactive maximum. Additionally, the randomly drawn MCMC samples appear to diverge from the best fit. The offset time found by P15 (-22.5 days) is different to the offset time found by every other model (about -21 days) which accounts for the smaller deviation. The larger deviation and the divergence from the randomly drawn MCMC samples can be explained by the non-Gaussian posterior of P15's $R_{e}$ parameter. This non-Gaussian profile means that the median of the posterior does not lie at the maximum of the posterior. P20 suffers a similar issue of non-Gaussianity, with the posterior of $M_{e}$ being multimodal (two peaks), although it does perform significantly better than P15 despite making similar assumptions. Unlike the SW17 models which have smaller residuals, neither the P15 nor P20 model make any assumption about the density of the progenitor which could account for their non-Gaussian posteriors.

The SW17 n=3/2 and SW17 n=3 models have a reduced chi-squared of 6.57 and 6.39, respectively; these are betters fits to the ground-based data than P15 and P20 which have a reduced chi squared of 17.79 and 12.82, respectively. All models have relatively large reduced chi squared (with a value of 1 being desirable). This could be due to the inherent systematics in the~\kep{} data, or could be indicative of some physics not paramaterised by these models.

When comparing the best fitting velocities to the velocity measured from the spectrum, 10,200 km s$^{-1}$, we see that the SW17 n=3/2 once again best matches the data with a best fitting velocity of $\sim$8,800 km s$^{-1}$, as opposed to $\sim$14,000 km s$^{-1}$, $\sim$18,000 km s$^{-1}$, and $\sim$8,100 km s$^{-1}$ for P15, P20, and SW17 n=3, respectively. As such we conclude that SW17 n=3/2 is the best model for SN 2017jgh.

Overall it seems that the SW17 models are more physically accurate than P15 and P20 owing to their density assumption with SW17 n=3/2 being the best fitting model for this supernova. This suggests that the progenitor of SN 2017jgh had an envelope radius of $\sim$130$R_{\odot}$ and an envelope mass of $\sim$0.50$M_{\odot}$. This radius is similar to the progenitor radius of SN 1993J~\citep{MaundSmartt2004}, and another well studied IIb supernova SN 2011dh ($\sim200_{\odot}$~\citet{BerstenBenvenuto2012}), both of which are believed to have yellow supergiant progenitors. We conclude that the progenitor of SN 2017jgh was likely a yellow supergiant. This is reinforced by the similarities SN 2017jgh had with SN 1993J, in both its lightcurve and colour evolution.

\section{The Importance of the Rise}\label{sec:rise}
\begin{figure*}
    \centering
    \includegraphics[width=\linewidth]{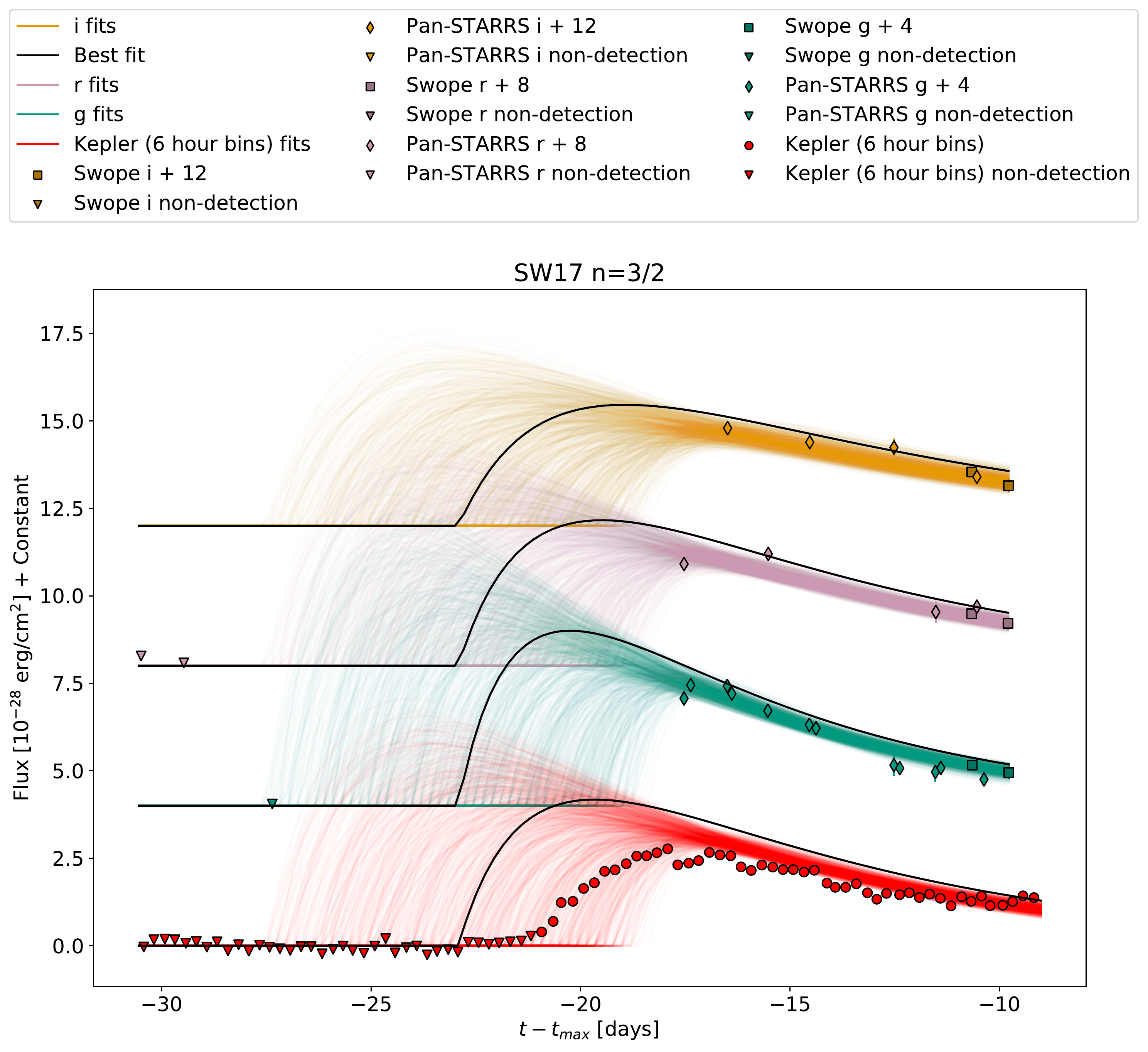}
    \caption{Fits to the ground-based lightcurve, ignoring all~\kep{} data. This emulates the conditions of most previous~\scc{} observations. The coloured lines are 1000 randomly drawn samples from the MCMC chain which give a visual for the shape of the posterior. The black line is the median value of the posterior. Note that the median is taken from each parameter posterior independently so will differ from the randomly drawn samples. It is obvious that these fits do not constrain the rise time, and even though they seem reasonably when compared to the ground-based data, when applied to the~\kep{} data we see that they are quite inaccurate.}
    \label{fig:RSG}
\end{figure*}
\begin{figure*}
    \centering
    \includegraphics[width=\linewidth]{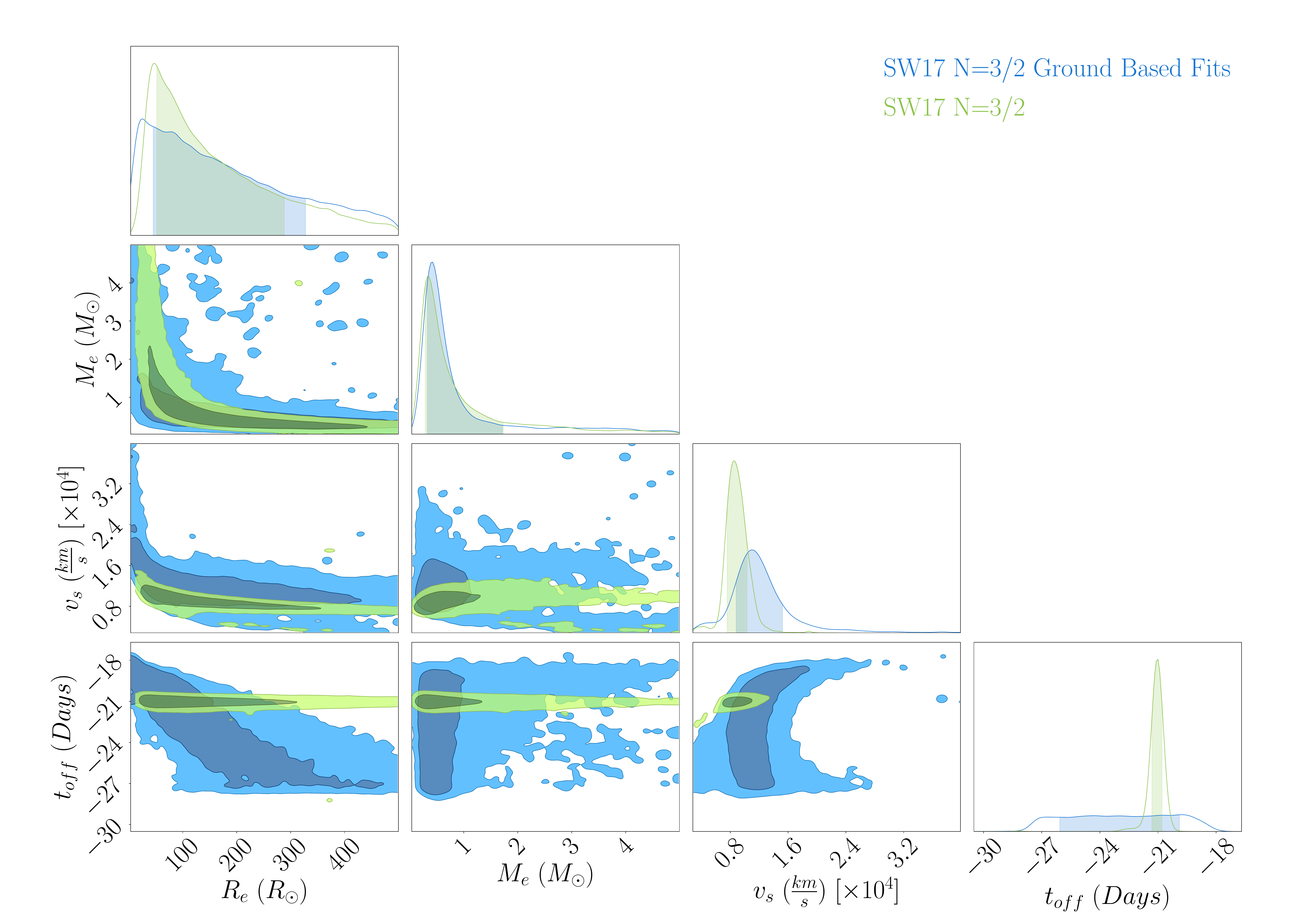}
    \caption{A comparison of the posteriors of the ground-based fits (blue) and the fits to the full lightcurve (green) for the SW17 n=3/2 model. The ground-based fits do not constrain the rise time and as such have a systematic offset in the $v_{s}$ parameter. However the ground-based fits do appear to constrain the $R_{e}$ and $M_{e}$ parameters at least as well as the full lightcurve fits.}
    \label{fig:groundCorner}
\end{figure*}
\begin{table*}
    \centering
    \begin{tabular}{c|ccccc}
         Model & $R_{e}~\left(R_{\odot}\right)$ & $M_{e}~\left(M_{\odot}\right)$ & $v~\left(\mathrm{km}/\mathrm{s}\right)$ & $t_{off}~\left(t-t_{max}\right)~(days)$ & $\chi^{2}$ \\
         SW17 n=3/2 & $150^{+180}_{-100}$ & $0.55^{+1.16}_{-0.23}$ & $\left(11.5^{+3.7}_{-2.7}\right)\times10^{3}$ & $-22.9^{+3.0}_{-3.1}$ & 8.66 \\
    \end{tabular}
    \caption{Best fitting parameters for fits to the ground-based lightcurve with the SW17 n=3/2 model, found by taking the 16th, 50th, and 84th percentile as the lower bound, best value, and upper bound, respectively. The reduced chi squared of the best fitting value is also included.}
    \label{tab:groundResults}
\end{table*}
As this is the first complete example of a high cadence shock cooling lightcurve, it provides an opportunity to evaluate how important the information contained in the rise is to getting an accurate fit. We perform fits to the ground-based lightcurve of SN 2017jgh without any~\kep{} data. The ground-based lightcurve contains no rise information and is quite similar in coverage and cadence to the lightcurves of SN 2016gkg and SN 1993J. Figure~\ref{fig:RSG} shows the results of these fits for the SW17 n=3/2 model, the model which best fit the full lightcurve. The best fitting values are given in Table~\ref{tab:groundResults}, and a comparison between the posteriors of the ground-based fits and the fits to the full lightcurve are shown in Figure~\ref{fig:groundCorner}.

Although the fits to the ground-based portion of the lightcurve appear reasonable, the extrapolated fits to the rise portion of the~\scc{} are very poorly constrained. Additionally, the best fit does not match the~\kep{} lightcurve. Overall, this suggests that the decline of the~\scc{} is not enough to constrain the rise of the~\scc{}.

The best fitting radius and mass of the ground-based fits ($\sim150~R_{\odot}$ and $\sim0.55~M_{\odot}$) are consistent with the radius and mass of the full lightcurve fits ($\sim130$ and $\sim0.56$), however both the velocity and offset time are significantly different at $\sim11,000$ km s$^{-1}$ and $\sim-23$ days compared to the full lightcurve fits of $\sim8,800$ km s$^{-1}$ and $\sim-21$ days. This suggests that the rise is required to constrain the velocity and explosion time, however the envelope radius and envelope mass can be constrained with just the decline.

In order to estimate the numerical impact of the~\scc{} rise on the quality of the fits, we calculate the percentage difference between the fits with and without the rise. We make use of bootstrap resampling to approximate the population mean and uncertainty. For both the full lightcurve fit and the decline fit we resample 10000 chains, each of which contain 500000 samples of the posterior. For each of these resampled chains we find the median of the parameters, and then calculate the mean and standard deviation of these medians. The mean, and percentage difference between the full lightcurve fit and the decline fit is show in table~\ref{tab:percent}. The envelope mass is the least affected, with only a 1\% difference between the fits. The envelope radius, offset time, and velocity are more greatly affected, at 15\%, 20\%, and 25\%, respectively. These values provide a rough approximation for the systematic uncertainty of not having the rise of the~\scc{}.
\begin{table*}
    \centering
    \begin{tabular}{l|cccc}
         & $R_{e}~(R_{\odot})$ & $M_{e}~(M_{\odot})$ & $v~(10^{3})$ km~s$^{-1}$ & $t_{off}~(t-t_{max})~(days)$  \\
        Full lightcurve fit & $127.82\pm0.21$ & $0.56\pm0.0008$ & $8.81\pm0.0023$ & $9.5\pm0.00024$ \\
        Decline fit & $149.11\pm0.25$ & $0.55\pm0.00055$ & $11.52\pm0.0047$ & $7.62\pm0.0062$ \\
        Percentage difference & $15.38\pm0.67$\% & $1.13\pm0.49$\% & $26.67\pm0.14$\% & $22\pm0.23$\%
    \end{tabular}
    \caption{Average parameter median over 10000 bootstrap resamples for both the full lightcurve fit and decline fit, as well as the percentage difference between them.}
    \label{tab:percent}
\end{table*}

\section{Conclusions}\label{sec:conclusions}
The high cadence~\kep{} lightcurve of SN 2017jgh provides a unique opportunity to investigate the complete shock cooling lightcurve of a Type IIb SNe. We fit the lightcurve with four models, the linearly expanding P15 model, the two-component P20 model, and the polytropic SW17 models with $n=3/2$ and $n=3$, modelling a red supergiant, and blue supergiant, respectively.

After fitting the P15, P20, SW17 $n=\nicefrac{3}{2}$ and SW17 $n=3$ models to SN 2017jgh, we found that the SW17 $n=\nicefrac{3}{2}$ model provides the best fit (with a reduced $\chi^{2}$ of 6.57). This fit suggests that the progenitor of SN 2017jgh was a yellow supergiant with an envelope radius of $\sim50-290~R_{\odot}$, an envelope mass of $\sim0-1.7~M_{\odot}$, a shock velocity of $\left(7.5-10.3\right)\times10^3$ km s$^{-1}$, and an offset time of $\sim-21$ days until radioactive maximum. SW17 $n=\nicefrac{3}{2}$ is also best able to reproduce the measured shock velocity of 9100$\pm$470 km s$^{-1}$.

Furthmore, we showed that the P15 and P20 models had difficulty reproducing the decline of the~\scc{}, overestimating and underestimating it, respectively. This is possibly due to the lack of density information in these models. By comparison, both the SW17 $n=3$ and SW17 $n=\nicefrac{3}{2}$ models follow the shape of the lightcurve better with very little deviation in the residual. The polytropic density model appears to be a better approximation for the true density of the shock cooling lightcurve.

In addition to determining the parameters of the progenitor of SN 2017jgh, we use the complete~\scc{} as an opportunity to investigate how important the rise of the shock cooling lightcurve is to the quality of fits. To do so, we fit the ground-based lightcurve of SN 2017jgh ignoring any~\kep{} data. These fits were unable to constrain the explosion time and could not recover the~\kep{} lightcurve. This lead to a systematic offset in $v_{s}$, but both $R_{e}$ and $M_{e}$ seem mostly unaffected. In all models velocity and explosion time are degenerate, and radius and mass are degenerate. It appears that the rise constrains the offset time and velocity whilst the decline constrains the radius and mass. The rise is more sensitive to temperature and density~\citep{Piro2015,SapirWaxman2017,PiroHaynie2020}, so the shock velocity will also be sensitive to the shape of the rise.

Overall this suggests that the cadence and rise information contained within the~\kep{} lightcurve is invaluable to getting a good fit, specifically to constraining the offset time and shock velocity. This is less important for constraining the envelope radius and mass. Long term, high cadence surveys akin to~\kep{} will be essential for improving these analytical~\scc{} models.
\section{Acknowledgments}
This is a pre-copyedited, author-produced PDF of an article accepted for publication in the Monthly Notice of the Royal Astronomical Society following peer review. The version of record~\citep{ArmstrongTucker2021} is available online at: \url{https://doi.org/10.1093/mnras/stab2138}.

P.A. and B.E.T. acknowledge parts of this research was carried out on the traditional lands of the Ngunnawal people.  We pay our respects to their elders past, present, and emerging.
This paper includes data collected by the K2 mission. Funding for the K2 mission is provided by the NASA Science Mission directorate. 
KEGS is supported in part by NASA K2 cycle 4, 5, and 6 grants NNX17AI64G and 80NSSC18K0302, and 80NSSC19K0112, respectively.
Pan-STARRS (PS1) is supported in part by the National Aeronautics and Space Administration under Grants NNX12AT65G and NNX14AM74G. The PanSTARRS1 Surveys (PS1) and the PS1 public science archive have been made possible through contributions by the Institute for Astronomy, the University of Hawaii, the Pan-STARRS Project Office, the Max-Planck Society and its participating institutes, the Max Planck Institute for Astronomy, Heidelberg and the Max Planck Institute for Extraterrestrial Physics, Garching, The Johns Hopkins University, Durham University, the University of Edinburgh, the Queen’s University Belfast, the Harvard-Smithsonian Center for Astrophysics, the Las Cumbres Observatory Global Telescope Network Incorporated, the National Central University of Taiwan, the Space Telescope Science Institute, the National Aeronautics and Space Administration under Grant No. NNX08AR22G issued through the Planetary Science Division of the NASA Science Mission Directorate, the National Science Foundation Grant No. AST–1238877, the University of Maryland, Eotvos Lorand University (ELTE), the Los Alamos National Laboratory, and the Gordon and Betty Moore Foundation
Based on observations at Cerro Tololo Inter-American Observatory, National Optical Astronomy Observatory (NOAO 2017B-0279; PI: A Rest, NOAO 2017B-0285; PI: A Rest), which is operated by the Association of Universities for Research in Astronomy (AURA) under a cooperative agreement with the National Science Foundation
Based on observations obtained at the international Gemini Observatory, which is managed by the Association of Universities for Research in Astronomy (AURA) under a cooperative agreement with the National Science Foundation on behalf of the Gemini Observatory partnership: the National Science Foundation (United States), National Research Council (Canada), Agencia Nacional de Investigaci\'{o}n y Desarrollo (Chile), Ministerio de Ciencia, Tecnolog\'{i}a e Innovaci\'{o}n (Argentina), Minist\'{e}rio da Ci$\hat{\text{e}}$ncia, Tecnologia, Inova\c{c}\~{o}es e Comunica\c{c}\~{o}es (Brazil), and Korea Astronomy and Space Science Institute (Republic of Korea). Observations in this program were obtained through program ID GS-2017B-LP-13.
P. A. was supported by an Australian Government Research Training Program (RTP) Scholarship.
B. E. T. and his group  were supported by the Australian Research Council Centre of Excellence for All Sky Astrophysics in 3 Dimensions (ASTRO 3D), through project number CE170100013.
The UCSC transient team is supported in part by NASA/{\it K2} grants 80NSSC18K0303 and 80NSSC19K0113, the Gordon \& Betty Moore Foundation, the Heising-Simons Foundation, and by a fellowship from the David and Lucile Packard Foundation to R.J.F.  D. O. J. acknowledges support provided by NASA Hubble Fellowship grant HST-HF2-51462.001, which is awarded by the Space Telescope Science Institute, operated by the Association of Universities for Research in Astronomy, Inc., for NASA, under contract NAS5-26555.
I. A. is a CIFAR Azrieli Global Scholar in the Gravity and the Extreme Universe Program and acknowledges support from that program, from the European Research Council (ERC) under the European Union’s Horizon 2020 research and innovation program (grant agreement number 852097), from the Israel Science Foundation (grant number 2752/19), from the United States - Israel Binational Science Foundation (BSF), and from the Israeli Council for Higher Education Alon Fellowship.
M. R. D. acknowledges support from the NSERC through grant RGPIN-2019-06186, the Canada Research Chairs Program, the Canadian Institute for Advanced Research (CIFAR), and the Dunlap Institute at the University of Toronto. 
D. A. C. acknowledges support from the National Science Foundation Graduate Research Fellowship under Grant DGE1339067.
This project has been supported by the LP2018-7 Lend\"{u}let grant of the Hungarian Academy of Sciences.
L. G. acknowledges financial support from the Spanish Ministry of Science, Innovation and Universities (MICIU) under the 2019 Ram\'on y Cajal program RYC2019-027683 and from the Spanish MICIU project PID2020-115253GA-I00.
B. J. S. is supported by NASA grant 80NSSC19K1717 and NSF grants AST-1920392 and AST-1911074.
Support for T. W. S. H was provided by NASA through the NASA Hubble Fellowship grant HST-HF2-51458.001-A awarded by the Space Telescope Science Institute, which is operated by the Association of Universities for Research in Astronomy, Inc., for NASA, under contract NAS5-26555.
Research by S.V. is supported by NSF grants AST-1813176 and AST-2008108
L. K. acknowledges the financial support of the Hungarian National Research, Development and Innovation Office grant NKFIH PD-134784.
L. K. and Zs. B. are supported by the J\`{a}nos Bolyai Research Scholarship of the Hungarian Academy of Sciences.
The Konkoly team has been supported by the project "Transient Astrophysical Objects"  GINOP 2.3.2-15-2016-00033  of the National Research, Development and Innovation Office (NKFIH), Hungary, funded by the European Union.
S. W. J. acknowledges support from US National Science Foundation award AST-1615455.
This research has made use of the SVO Filter Profile Service (http://svo2.cab.inta-csic.es/theory/fps/) supported from the Spanish MINECO through grant AYA2017-84089.
The LCO team is supported by NASA grant 80NSSC19K0119 and NSF grants AST-1911225 and AST-1911151.
\subsection{Data availability}
The data underlying this article are available in the article and in its online supplementary material.
\subsection{Affiliations\label{sec:affiliation}}
\textit{\small{
$^{1}$Mt Stromlo Observatory, The Research School of Astronomy and Astrophysics, Australian National University, ACT 2601, Australia\\
$^{2}$National Centre for the Public Awareness of Science, Australian National University, Canberra, ACT 2611, Australia\\
$^{3}$The ARC Centre of Excellence for All-Sky Astrophysics in 3 Dimensions (ASTRO 3D)\\
$^{4}$Space Telescope Science Institute, 3700 San Martin Dr., Baltimore, MD 21218, USA\\
$^{5}$Department of Physics and Astronomy, The Johns Hopkins University, 3400 North Charles Street, Baltimore, MD 21218, USA.\\
$^{6}$The Observatories of the Carnegie Institution for Science, 813 Santa Barbara St., Pasadena, CA 91101, USA\\
$^{7}$School of Mathematics and Physics, University of Queensland, Brisbane, QLD 4072, Australia\\
$^{8}$Australian Astronomical Observatory, North Ryde, NSW 2113, Australia\\
$^{9}$Gemini Observatory/NSF’s NOIRLab, Casilla 603, La Serena, Chile\\
$^{10}$University of Illinois at Urbana-Champaign, 1003 W. Green St., IL 61801, USA\\
$^{11}$Centre for Astrophysical Surveys, National Centre for Supercomputing Applications, Urbana, IL 61801, USA\\
$^{12}$Astronomy Department, University of Maryland, College Park, MD 20742-2421, USA\\
$^{13}$The School of Physics and Astronomy, Tel Aviv University, Tel Aviv 69978, Israel\\
$^{14}$CIFAR Azrieli Global Scholars program, CIFAR, Toronto, Canada\\
$^{15}$David A. Dunlap Department of Astronomy and Astrophysics, University of Toronto, 50 St. George Street, Toronto, Ontario, M5S 3H4 Canada\\
$^{16}$Department of Astronomy and Astrophysics, University of California, Santa Cruz, CA 95064, USA\\
$^{17}$Department of Astronomy, University of Texas at Austin, Austin, TX, 78712, USA\\
$^{18}$Centro de Astronom\'{i}a (CITEVA), Universidad de Antofagasta, Avenida U. de Antofagasta 02800, Antofagasta, Chile\\
$^{19}$Las Campanas Observatory, Carnegie Observatories, Casilla 601, La Serena, Chile\\
$^{20}$Max-Planck-Institut fur Astrophysik, Karl-Schwarzschild-Str 1, D-85748 Garching bei M\"{u}nchen, Germany \\
$^{21}$Center for Interdisciplinary Exploration and Research in Astrophysics (CIERA)\\
$^{22}$Department of Physics and Astronomy, Northwestern University, Evanston, IL 60208, USA\\
$^{23}$Centre for Astrophysics Research, School of Physics, Astronomy and Mathematics, University of Hertfordshire, College Lane, Hatfield AL10 9AB, UK\\
$^{24}$University of Notre Dame, Notre Dame, IN 46556, USA\\
$^{25}$Institute for Astronomy, University of Hawai\`{}i at Manoa, 2680 Woodlawn Drive, Honolulu, HI 96822, USA\\
$^{26}$Exoplanets  and  Stellar  Astrophysics  Laboratory,  Code  667,NASA Goddard Space Flight Center, Greenbelt, MD 20771, USA\\
$^{27}$University of Maryland, Baltimore County, 1000 Hilltop Circle, Baltimore, MD 21250, USA\\
$^{28}$Bay Area Environmental Research Institute, P.O. Box 25, Moffett Field, CA 94035, USA\\
$^{29}$NASA Ames Research Center, Moffett Field, CA 94035, USA\\
$^{30}$Astrophysics Research Centre, School of Mathematics and Physics,Queen’s University Belfast, BT7 1NN, UK\\
$^{31}$Department of Astronomy and Theoretical Astrophysics Center, University of California, Berkeley, CA 94720, USA\\
$^{32}$Department of Physics, University of California, Berkeley, CA 94720, USA\\
$^{33}$Lawrence Berkeley National Laboratory, Berkeley, CA, USA\\
$^{34}$Konkoly Observatory, Research Centre for Astronomy and Earth Sciences, E\"{o}tv\"{o}s Lor\`{a}nd Research Network (ELKH), H-1121 Konkoly Thege Mikl\`{o}s \`{u}t 15-17, Budapest, Hungary\\
$^{35}$MTA CSFK Lend\"{u}let Near-Field Cosmology Research Group\\
$^{36}$Coral Towers Observatory, Cairns, QLD 4870, Australia\\
$^{37}$Institute of Space Sciences (ICS, CSIC), Campus UAB, Carrer de Can Magrans, s/n, E-08193 Barcelona, Spain\\
$^{38}$Department of Physics, University of California, Santa Barbara, CA 93106-9530, USA\\
$^{39}$Las Cumbres Observatory, 6740 Cortona Dr, Suite 102, Goleta, CA 93117-5575, USA\\
$^{40}$Institute of Physics, University of Szeged, Dom ter 9, Szeged, 6720 Hungary\\
$^{41}$Departamento de Astronom\'{\i}a y Astrof\'{\i}sica, Universidad de Valencia, E-46100 Burjassot, Valencia, Spain\\
$^{42}$Observatorio Astron\'omico, Universidad de Valencia, E-46980 Paterna, Valencia, Spain\\
$^{43}$Graduate Institute of Astronomy, National Central University, 300 Jhongda Road, 32001 Jhongli, Taiwan\\
$^{44}$Physics Department and Tsinghua Center for Astrophysics (THCA), Tsinghua University, Beijing, 100084, China\\
$^{45}$Steward Observatory, University of Arizona, 933 North Cherry Avenue, Tucson, AZ 85721-0065, USA\\
$^{46}$ELTE E\"{o}tv\"{o}s Lor\'{a}nd University, Institute of Physics, Budapest, Hungary\\
$^{47}$Department of Astronomy, Columbia University, New York, NY 10027-6601, USA\\
$^{48}$Beijing Planetarium, Beijing Academy of Science and Technology,Beijing, 100044\\
$^{49}$Physics Department, Tsinghua University, Beijing, 100084\\
$^{50}$Cerro Tololo Inter-American Observatory, NSF’s National Optical-Infrared Astronomy Research Laboratory, Casilla 603, La Serena, Chile\\
$^{51}$Department  of  Physics  and  Astronomy,  Texas  A\&M University, 4242 TAMU, College Station, TX 77843, USA\\
$^{52}$Department of Mathematics, University of York, Heslington, York, YO10 5DD, United Kingdom\\
$^{53}$Department of Physics and Astronomy, University of California, 1 Shields Avenue, Davis, CA 95616-5270, USA\\
$^{54}$European Southern Observatory, Karl-Schwarzschild-Strasse 2, 85748 Garching bei M \"{u}nchen, German\\
$^{55}$Department of Physics and Astronomy, Rutgers the State University of New Jersey, 136 Frelinghuysen Road, Piscataway, NJ 08854, USA\\
$^{56}$Centre for Gravitational Astrophysics, College of Science, The Australian National University, ACT 2601, Australia\\
$^{57}$School of Physics, The University of Melbourne, Parkville, VIC 3010, Australia
}}\\
\subsection{Software}
NumPy~\citep{numpy}, MatPlotLib~\citep{matplotlib}, SciPy~\citep{scipy}, AstroPy~\citep{astropy}, emcee~\citep{ForemanMackeyFarr2019}, scikit-learn~\citep{scikit-learn}, chainconsumer~\citep{chainconsumer}
\subsection{Facilities}
Kepler~\citep{HowellSobeck2014}, Gemini~\citep{HookJorgensen2004}, Swope~\citep{FolatelliPhillips2010}, PS1~\citep{ChambersMagnier2016}
\pagebreak
\bibliographystyle{mnras}
\bibliography{paper.bib}
\bsp
\label{lastpage}
\end{document}